# Moiré Engineering of Cooper-Pair Density Modulation States

Zihao Wang[1,6], Bing Xia[1,6], Stephen Paolini[1,6], Zi-Jie Yan[1], Pu Xiao[1], Jiatao Song[1], Veer Gowda[1], Hongtao Rong[1], Di Xiao[2,3], Xiaodong Xu[2,3], Weida Wu[4], Ziqiang Wang[5], and Cui-Zu Chang[1]

[1]Department of Physics, The Pennsylvania State University, University Park, PA 16802, USA

[2]Department of Physics, University of Washington, Seattle, WA 98195, USA.

[3]Department of Materials Science and Engineering, University of Washington, Seattle, WA 98195, USA.

[4]Department of Physics and Astronomy, Rutgers University, Piscataway, NJ 08854, USA

[5]Department of Physics, Boston College, Chestnut Hill, MA, 02467, USA

[6]These authors contributed equally: Zihao Wang, Bing Xia, and Stephen Paolini

Corresponding author: cxc955@psu.edu (C.-Z. C.).

**Abstract: Cooper-pair density modulation (CPDM) states are superconducting phases in which the order parameter varies periodically in real space without breaking translational symmetry[1-3]. Recently, moiré superlattices in layered materials[4-20] have emerged as powerful platforms for engineering charge density with tunable lattice symmetry, offering a new route to creating and controlling CPDM states. In this work, we demonstrate moiré-induced CPDM states in a bilayer heterostructure formed by epitaxially stacking one quintuple layer (1 QL) of topological insulator $Sb_2Te_3$ on a six-unit-cell (6 UC) antiferromagnetic FeTe layer. Scanning tunneling microscopy and spectroscopy (STM/S) measurements reveal a moiré superlattice formed between the hexagonal Te lattice of $Sb_2Te_3$ and the square Te lattice of FeTe, which spatially modulates the two superconducting gaps of the 1 QL $Sb_2Te_3$/6 UC FeTe bilayer. Our Josephson STM/S measurements provide direct real-space imaging of the**

**CPDM states with a wavelength corresponding to the periodicity of the moiré superlattice. By substituting $Sb_2Te_3$ with $Bi_2Te_3$, we achieve control over both the periodicity and magnitude of the CPDM states. Our work demonstrates an epitaxial strategy for synthesizing moiré superlattices from materials with different crystal symmetries and reveals a new mechanism for engineering CPDM states in designer bilayer heterostructures.**

**Main text:** More than half a century ago, it was proposed that weak-coupling spin-singlet superconductors under strong Zeeman magnetic fields can host finite-momentum Cooper pairing, leading to phase-modulated[21] or amplitude-modulated[22] order parameters, collectively known as Fulde-Ferrell-Larkin-Ovchinnikov (FFLO) states. Unlike FFLO states, Cooper-pair density wave (PDW) states emerge *spontaneously* from strong interactions without breaking time-reversal symmetry. PDW states have attracted significant interest due to their potential as a phase-incoherent state in the pseudogap regime of cuprates[23-25], where they have been proposed as a key to understanding both the high superconducting transition temperature and unconventional pairing mechanisms. The PDW order breaks lattice translational symmetry and can emerge either as a primary order or a subsidiary order of a pre-existing density wave order. A real-space inter-unit-cell modulation of the pairing gap, which is a hallmark of the PDW state, has been observed in a variety of materials, including cuprates[23-31], iron-based superconductors[32,33], transition metal dichalcogenides[34,35], kagome metals[36-38], and $UTe_2$(ref.[39]). More recently, Cooper-pair density modulation (CPDM) states have been reported in iron-based superconductors, featuring intra-unit-cell modulation of the superconducting order parameter without breaking lattice translational symmetry[1-3,40]. However, to date, the periodicity of CPDM states has been dictated by the crystal lattice, thus limiting external tunability and hindering systematic control over the superconducting order parameter.

Over the past decade, synthetic moiré materials have opened new avenues for engineering tunable periodic potentials. When two atomically thin layers are stacked together with a twist angle or lattice mismatch, a moiré superlattice potential forms. In moiré materials, many interesting phenomena have been discovered, including unconventional superconductivity[4,6,8-10], correlated insulators[5,11], and the quantum anomalous Hall effect[7,12-14]. Meanwhile, STM/S measurements have directly visualized generalized Wigner crystals[15-17], layer-pseudospin skyrmions[18,19], and twist-dependent Bogoliubov quasiparticle interference (QPI)(ref.[20]) in moiré materials. Despite these advances, nearly all moiré superlattices studied to date have been created by stacking layered materials with identical crystal symmetry[4-20]. Moiré patterns formed between layers of different crystal symmetries, as well as moiré-induced CPDM states, remain unexplored. A platform in which the periodicity and magnitude of CPDM states can be designed and tailored would enable systematic studies of the interplay between superconductivity and a moiré superlattice potential.

In this work, we employ molecular beam epitaxy (MBE) to synthesize superconducting bilayers formed by stacking a single quintuple layer (QL) of topological insulator $Sb_2Te_3$ on a 6 unit cell (UC) antiferromagnetic FeTe film. The lattice mismatch between the hexagonal Te layer of 1QL $Sb_2Te_3$ and the square Te layer of 6UC FeTe generates a rhombic moiré superlattice (Fig. 1). STM/S measurements on the 1QL $Sb_2Te_3$ layer reveal two superconducting gaps that are periodically modulated by the moiré superlattice. Josephson STM/S measurements directly visualize the CPDM states in 1QL $Sb_2Te_3$/6UC FeTe bilayers. Furthermore, replacing $Sb_2Te_3$ with $Bi_2Te_3$ leads to changes in both CPDM periodicity and magnitude. Our work demonstrates a distinct Cooper pairing state originating from moiré-superlattice-induced charge density modulation (CDM) and establishes engineered moiré superlattices as a new platform to investigate the interplay between superconductivity and a large and tunable real-space length scale.

**Moiré pattern in 1QL Sb$_2$Te$_3$/6UC FeTe**

Figure 1a,b shows the top views of the tetragonal FeTe and trigonal Sb$_2$Te$_3$ lattice structures, respectively. The in-plane lattice constant $a_1$ of FeTe is ~3.82Å(ref.[41]), while that of Sb$_2$Te$_3$ $a_2$ is ~4.27Å(ref.[42]). A schematic of the STM/S measurement setup for 1QL Sb$_2$Te$_3$/6UC FeTe bilayers is shown in Fig. 1c. The bottom 6UC FeTe layer exhibits large atomically flat terraces with square pyramidal structures (Extended Data Fig. 1). In comparison, the top 1QL Sb$_2$Te$_3$ layer forms islands with a typical size of ~50nm (Fig. 1d). An atomic resolution STM image shows both a hexagonal Te lattice of Sb$_2$Te$_3$ and a moiré pattern (Fig. 1e). In the corresponding Fourier transform (FT) image, 6 Bragg peaks corresponding to the hexagonal Te lattice of Sb$_2$Te$_3$ and 4 peaks associated with the moiré superlattice are observed (Fig. 1f). The 4 moiré wavevectors $\mathbf{Q}_{\text{moiré}}$ form a rectangle and can be grouped into two pairs ($\pm\mathbf{Q}_1$ and $\pm\mathbf{Q}_2$). The FT-filtered image at $\mathbf{Q}_{\text{moiré}}$ shows a rhombic shape in real space (Fig. 1h), which can be understood by examining the Te sublattices of the Sb$_2$Te$_3$ and FeTe layers (Fig. 1g). By analyzing the positions of the moiré superlattice peaks (Fig. S10), we determine that the moiré superlattice has a periodicity of ~$\sqrt{3}a_2\times 11a_2$ along the minor and major diagonals, respectively. The temperature dependence of the average tunneling spectra $g=\mathrm{d}I/\mathrm{d}V$ on 1QL Sb$_2$Te$_3$ shows that the superconducting gap closes at $T$~12K (Fig. 1i), corresponding to the superconducting transition temperature $T_c$, consistent with the previously reported $T_c$ values for Sb$_2$Te$_3$/FeTe bilayers[43,44].

**Moiré-induced CPDM states**

We first investigate how the moiré superlattice modulates superconductivity in 1QL Sb$_2$Te$_3$/6UC FeTe bilayers (Fig. 2a). Figure 2b shows typical d$I$/d$V$ spectra. The d$I$/d$V$ spectra near zero bias exhibit a U-shape feature with two pairs of coherence peaks. Although the mechanism behind the emergent interface-induced superconductivity in FeTe-based heterostructures remains unclear[43-47], the spectral features on 1QL $Sb_2Te_3$ resemble those observed in multiband superconductors such as Fe(Se,Te) and FeSe(refs.[48-53]). The two spectra in Fig. 2b reveal that the moiré pattern modulates both the superconducting gaps and the spectral weights of the coherence peaks. To better visualize this modulation, we fit the spectra using the Dynes formula[54,55] with two isotropic *s*-wave superconducting gaps to extract $\Delta_{1,2}$ (Extended Data Fig. 2a).

Next, we analyze a spectral linecut along the minor diagonal direction (green arrow, Fig. 2a) of the rhombic moiré pattern (Fig. 2c,d). Figure 2e summarizes the height profile $Z$ and the extracted $\Delta_{1,2}$. $Z$, simultaneously measured with the d$I$/d$V$ spectra at $V_{bias}$=10mV, shows good consistency with the average $Z$ extracted from the atomic resolution STM image (Extended Data Fig. 3). Both $\Delta_1$ and $\Delta_2$ show a local minimum at Te atom sites (Fig. 2e), while exhibiting two local maxima: one located between Te- and Te+ from two adjacent moiré unit cells, and another between Te- and Te+ within the same moiré unit cell. This behavior indicates a phase shift between the $\Delta_{1,2}$ map and the moiré superlattice. The spectral linecut along the major diagonal direction (red arrow, Fig. 2a) also exhibits clear modulation of $\Delta_{1,2}$ (Extended Data Fig. 2c,d). As noted above, the rhombic moiré superlattice is not aligned with the high symmetry directions of the hexagonal Te lattice in the 1QL $Sb_2Te_3$ layer. Consequently, the sharp peaks at $q=2\pi/(\sqrt{3}a_2)$ and $2q=4\pi/(\sqrt{3}a_2)$ in 1D FT of $\Delta_{1,2}$ correspond to projections of the wavevectors in the 2D FT images onto the minor diagonal direction in Fourier space (Fig. 2f). By applying a 1D lock-in method at $q$, we quantify a phase shift close to π between the $\Delta_{1,2}$ map and the moiré superlattice (Fig. 2g).

To fully characterize the 2D modulations of $\Delta_{1,2}$, we perform spectroscopic imaging-STM measurements over an 84×84 grid within a 14×14nm² area (Fig. 2a and Extended Data Fig. 4a). Figure 2h,i shows the spatial $\Delta_{1,2}$ maps, whose distributions provide the mean values of $\Delta_1$~2.58meV and $\Delta_2$~3.60meV (Fig. 2j). The FT images of $\Delta_{1,2}$ maps show 4 CPDM peaks at ±$\mathbf{Q}_1$ and $\pm \mathbf{Q}_2$ (Fig. 2k,l), which coincide with the moiré superlattice wavevectors $\mathbf{Q}_{\text{moiré}}$. This correspondence confirms that the $\Delta_{1,2}$ modulations conform to the moiré pattern in real space (Extended Data Fig. 4a). Among the 6 Bragg peaks, 4 peaks (blue circles) exhibit weak intensity, while the remaining 2 peaks (red circles) appear much stronger due to their overlap with the higher-order harmonics of the CPDM wavevectors. Therefore, the Te lattice of $Sb_2Te_3$ has a negligible contribution to the $\Delta_{1,2}$ modulation. In the 1D FT curves, the peaks at $q$ correspond to the projections of $\mathbf{Q}_1$ and $\mathbf{Q}_2$, while the peaks at $2q$ arise from the higher-order harmonic $\mathbf{Q}_1+\mathbf{Q}_2$ of the bidirectional CPDM states (Fig. 2f).

For visual clarity of $\Delta_{1,2}$ modulations, the $\Delta_{1,2}$ maps are filtered at $\pm \mathbf{Q}_1$ and $\pm \mathbf{Q}_2$ (Fig. 2m,n, and Extended Data Fig. 4f-i). The filtered $\Delta_{1,2}$ modulations exhibit apparent phase shifts relative to the filtered modulation of the moiré superlattice (Extended Data Fig. 4c-e). We employ a 2D lock-in method[56] to extract the phase information of the gap modulations $\phi_{\mathbf{Q}_{1,2}}^{\Delta_{1,2}}$ and moiré superlattice $\phi_{\mathbf{Q}_{1,2}}^{T}$ (Fig. S3). A phase shift close to $\pi$ between the gap modulations and the moiré superlattice is observed (Fig. 2o-p), where the phase shift is defined as $\delta\phi_{\mathbf{Q}_{1,2}}^{\Delta_{1,2}} = \phi_{\mathbf{Q}_{1,2}}^{\Delta_{1,2}} - \phi_{\mathbf{Q}_{1,2}}^{T}$. These phase shifts account for the $\Delta_{1,2}$ modulation in the 1D spectral linecut (Fig. 2e,g), providing strong evidence for intra-moiré-unit-cell CPDM states in 1QL $Sb_2Te_3$/6UC FeTe bilayers.

**Real-space imaging of the CPDM states**

To directly visualize the CPDM states, we perform Josephson STM/S measurements on the 1QL $Sb_2Te_3$/6UC FeTe bilayer using a superconducting Nb tip. In a phase-diffusive Josephson junction, the superfluid density $N_J(\mathbf{r})$ is proportional to $g_J(\mathbf{r},V=0\text{mV}) \times R_N^2(\mathbf{r})$, where $g_J(\mathbf{r},V=0\text{mV})$ is the Josephson peak height, and $R_N(\mathbf{r})$ is the normal state junction resistance[34,36,37,57]. Figure 3a shows an atomic resolution STM image measured using a Nb tip. Along the minor diagonal direction (red arrow, Fig. 3a), we measure linecut d$I$/d$V$ spectra with $R_N(\mathbf{r})$=2.5MΩ. Each d$I$/d$V$ spectrum reflects the convolution of the density of states (DOS) of the Nb tip and the local DOS of the sample. Three pairs of peaks are resolved in a typical d$I$/d$V$ spectrum (Fig. 3b): one corresponds to the coherence peaks of the Nb tip $\Delta_{tip}$, while the other two are located at $\Delta_{tip}+\Delta_1$ and $\Delta_{tip}+\Delta_2$. The spatial modulation of these convoluted conductance peaks conforms to the same periodicity as the moiré pattern (Fig. 3c,d). Through a numerical deconvolution procedure, we remove the contribution of the Nb tip and extract the local DOS of the sample (Extended Data Fig. 5). The modulation observed in the deconvoluted d$I$/d$V$ spectra (Extended Data Fig. 5c) is consistent with that obtained using a PtIr tip (Fig. 2c,d).

Next, we focus on the Josephson tunneling signals around $V_{bias}$=0mV. By gradually reducing $R_N$ to ~0.92MΩ, we observe a pronounced zero-bias peak in the d$I$/d$V$ spectra (Fig. 3e,g) and a double kink feature in the corresponding $I$-$V_{bias}$ curves (Fig. 3f), indicating the occurrence of Cooper-pair tunneling current. Along the minor diagonal direction (red arrow, Fig. 3a), we measure linecut d$I$/d$V$ spectra with reduced $R_N(\mathbf{r})$ and find that $g_J(\mathbf{r},V=0mV)$ exhibits spatial modulations consistent with the moiré pattern (Extended Data Fig. 5d,e). The color plot of $g_J(\mathbf{r},V)\times R_N^2(\mathbf{r})$ around $V_{bias}$=0mV shows a strong periodic modulation (Fig. 3h), providing a real-space imaging of the CPDM states in 1QL Sb$_2$Te$_3$/6UC FeTe bilayers. By comparing $N_J(\mathbf{r})$ with the simultaneously measured $Z$, we find that $N_J(\mathbf{r})$ reaches a maximum at Te$_+$ sites and a minimum between Te$_-$ and Te$_+$ sites across adjacent moiré unit cells (Fig. 3i). Although the FT of $N_J(\mathbf{r})$ (Fig. 3j) is consistent with the results obtained using a PtIr tip (Fig. 2f), $N_J(\mathbf{r})$ exhibits a nearly zero phase shift relative to the moiré superlattice (Fig. 3k), showing an anticorrelation with $\Delta_{1,2}$ modulations (Fig. 2g). We perform a Ginsburg-Landau free energy analysis to elucidate the emergence of CPDM states in 1 QL Sb$_2$Te$_3$/6UC FeTe bilayers and to clarify the interrelation among $Z$, $\Delta_{1,2}$, and $N_J(\mathbf{r})$ (Supplementary Information). We note that a similar anticorrelation has been reported in cuprates[58,59], suggesting that the CPDM states in 1QL Sb$_2$Te$_3$/6UC FeTe bilayers may host an unconventional Cooper-pair wavefunction.

**Moiré engineering of the CPDM states**

By tuning the Bi/Sb ratio, the lattice constant of the (Bi,Sb)$_2$Te$_3$ layer can be adjusted[43-45], thereby allowing programmable moiré superlattice potential and the associated CPDM states in 1QL (Bi,Sb)$_2$Te$_3$/6UC FeTe bilayers. To investigate the moiré engineering of CPDM states, we examine their manifestation in 1QL Bi$_2$Te$_3$/6UC FeTe bilayers. Figure 4a shows an atomic resolution STM image of an 1QL Bi$_2$Te$_3$/6UC FeTe bilayer, where both the hexagonal Te lattice and a moiré pattern are observed, resembling those in the 1QL Sb$_2$Te$_3$/6UC FeTe bilayer (Fig. 1e). Owing to the ~3% larger in-plane lattice constant of Bi$_2$Te$_3$ $a_3$~4.38Å compared to Sb$_2$Te$_3$(ref.[42]), the wavevector of the resulting moiré pattern is altered. The moiré superlattice in the 1QL Bi$_2$Te$_3$/6UC FeTe bilayer exhibits a periodicity of ~$\sqrt{3}a_3 \times 8a_3$ along the minor and major diagonals, respectively (Fig. 4b). A typical d$I$/d$V$ spectrum exhibits a U-shape superconducting gap $\Delta_1$ (Fig. 4c), which closes at $T$~12K (Extended Data Fig. 6a), consistent with the $T_c$ value determined from electrical transport measurements on Bi$_2$Te$_3$/FeTe bilayers[44,45].

Besides the coherence peaks, several peak features appear outside the superconducting gap $\Delta_1$, which also vanish at $T$~12K. Spectroscopic imaging STM measurements over $V_{bias}$ ranging from -8mV to 8mV show pronounced QPI patterns, as confirmed by the corresponding FT images (Fig. S8). Spectral linecut reveals that the peaks outside $\Delta_1$ exhibit strong QPI-induced spatial variation (Extended Data Fig. 6b). The QPI contribution can be suppressed by averaging the d$I$/d$V$ spectra, yielding a spatially averaged d$I$/d$V$ spectrum with two gap features, with $\Delta_1$~1.96meV and $\Delta_2$~3.32meV (Extended Data Fig. 7a). Figure 4d shows a typical d$I$/d$V$ spectrum measured with a Nb tip, where the coherence peaks of the Nb tip $\Delta_{tip}$ and convoluted conductance peaks $\Delta_{tip}+\Delta_1$ are observed. In the spatially averaged d$I$/d$V$ spectrum, the $\Delta_{tip}+\Delta_2$ peak is also resolved (Extended Data Fig. 7b). After removing the Nb tip contribution, the deconvoluted local DOS agrees well with the averaged d$I$/d$V$ spectrum measured with a PtIr tip (Extended Data Fig. 7a,c), further supporting the presence of multiband superconductivity in 1QL $Bi_2Te_3$/6UC FeTe bilayers. Moreover, the d$I$/d$V$ map $g[\mathbf{r},V\sim(\Delta_1+\Delta_{tip})/e]$, measured with a Nb tip, reveals that the amplitude of the convoluted coherence peak exhibits enhanced spatial modulation that conforms to the moiré superlattice in 1QL $Bi_2Te_3$/6UC FeTe bilayers (Extended Data Fig. 7d,e).

Next, we perform Josephson STM/S measurements on the 1QL $Bi_2Te_3$/6UC FeTe bilayer using a Nb tip to visualize the CPDM states. By mapping $R_N(\mathbf{r})$ (Extended Data Fig. 8d) and $g_J(\mathbf{r},V=0mV)$ (Extended Data Fig. 8c), we achieve the $N_J(\mathbf{r})$ map (Fig. 4e). The FT of $N_J(\mathbf{r})$ (Fig. 4f) closely matches that of the atomic resolution STM image (Fig. 4b), confirming that the CPDM states share the same periodicity as the moiré superlattice. The FT-filtered image at $\pm\mathbf{Q}_1$ and $\pm\mathbf{Q}_2$ clearly reveals the spatial modulation of $N_J(\mathbf{r})$ (Fig. 4g). Along the minor diagonal direction (red arrows, Fig. 4e,g), linecut d$I$/d$V$ spectra with reduced $R_N(\mathbf{r})$ are measured. The resulting color plot of $g_J(\mathbf{r},V)\times R_N(\mathbf{r})^2$ near $V_{bias}=0mV$ exhibits a periodic modulation (Fig. 4h). The normalized $N_J(\mathbf{r})$ modulation amplitudes suggest that the CPDM states in 1QL $Bi_2Te_3$/6UC FeTe bilayers are weaker than those in 1QL $Sb_2Te_3$/6UC FeTe bilayers (Fig. 4i), demonstrating the role of moiré engineering in tuning CPDM states in 1QL $(Bi,Sb)_2Te_3$/6UC FeTe bilayers.

**Discussion and outlook**

For 1QL $(Bi,Sb)_2Te_3$/6UC FeTe bilayers, the moiré superlattice potential can be written as $V(\vec{r}) = \sum_\alpha V_{\vec{Q}_\alpha} e^{i\vec{Q}_\alpha \cdot \vec{r}}$ ( $\vec{Q}_\alpha = \pm\vec{Q}_1, \pm\vec{Q}_2$ ). The moiré-superlattice-induced CDM is directly visualized in the d$I$/d$V$ maps $g(\mathbf{r},\pm100mV)$ of the 1QL $Sb_2Te_3$/6UC FeTe bilayers (Extended Data

Fig. 9c-e). Moreover, the coupling between CDM and the superconducting order parameter generates an induced CPDM state, independent of material-specific details. The amplitude of the CPDM state is proportional to the CDM strength and increases with the ratio $\lambda/\xi$, where $\lambda$ is the moiré unit cell size and $\xi$ is the superconducting coherence length (Supplementary Information). In moiré superconductors, $\lambda$ is highly tunable. When $\lambda$ is tuned to be comparable to or much larger than $\xi$, a CPDM state that follows the moiré superlattice can emerge (Fig. S16a). Therefore, CPDM states are expected to be a universal phenomenon in moiré superconductors, including twisted superconducting layers[60], twisted nonsuperconducting layers[4,6,8-10], and even moiré structures that combine superconducting and nonsuperconducting layers[20].

To summarize, we observe intra-moiré-unit-cell CPDM states in MBE-grown 1QL $(Bi,Sb)_2Te_3$/6UC FeTe bilayers. The superconducting gaps observed in these bilayers exhibit a two-gap feature, indicating multiband superconductivity in FeTe-based heterostructures[43-47]. By analyzing the modulation of the superconducting gap and Josephson tunneling signals, we demonstrate moiré engineering of CPDM states in 1QL $(Bi,Sb)_2Te_3$/6UC FeTe bilayers. These findings provide a new route for investigating the Cooper-pair symmetry of emergent superconductivity in FeTe-based heterostructures[43-47] and create a tunable and atomically precise platform for exploring how engineered moiré superlattice potentials influence superconductivity. Since thick $(Bi,Sb)_2Te_3$ films are well-established topological insulators[61-63], thick $(Bi,Sb)_2Te_3$/FeTe bilayers host superconductivity, Dirac surface states, and a moiré superlattice potential, making them a promising platform for moiré-enabled topological superconductivity[64]. Moreover, the $(Bi,Sb)_2Te_3$/FeTe bilayers also provide a platform to investigate the origin and Cooper pairing mechanism of moiré superconductivity.

**Methods**

**MBE growth**

Both 1 QL $Sb_2Te_3$/6 UC FeTe and 1 QL $Bi_2Te_3$/6 UC FeTe bilayers used in this work are grown in a commercial MBE chamber (Unisoku) with a base vacuum better than $2 \times 10^{-10}$ mbar. Metallic 0.05% Nb-doped $SrTiO_3$(100) substrates are used. Before MBE growth, the metallic $SrTiO_3$(100) substrates are first soaked in hot deionized water (~80 °C) for 2 hours, then immersed in ~4.5% HCl solution for 2 hours, and finally annealed at ~974 °C for 3 hours in a tube furnace with flowing oxygen. These treatments passivate and reconstruct the $SrTiO_3$(100) surface, making it suitable for the MBE growth of FeTe films. These heat-treated metallic $SrTiO_3$(100) substrates are loaded into the MBE chamber and outgassed at ~600 °C for 1 hour before the MBE growth. High-purity Fe (99.995%), Bi (99.9999%), Sb (99.9999%), and Te (99.9999%) are co-evaporated from Knudsen effusion cells. During MBE growth, the substrate temperature is maintained at ~330 °C for the bottom 6 UC FeTe layer and then reduced to ~240 °C for the top 1 QL $Sb_2Te_3$ or 1 QL $Bi_2Te_3$ layer. The growth rates are ~0.2 UC/min for FeTe, ~0.1 QL/min for $Sb_2Te_3$, and ~0.1 QL/min for $Bi_2Te_3$.

**STM/S measurements**

The STM/S measurements are performed in a Unisoku 1300 system with a base vacuum better than $2 \times 10^{-10}$ mbar. The system incorporates a single-shot $^3$He cryostat to achieve a base temperature of ~310 mK. The maximum magnetic field is ~11 T. Polycrystalline PtIr tips are used in our STM/S measurements. Before STM/S measurements on 1 QL $Sb_2Te_3$/6 UC FeTe or 1 QL $Bi_2Te_3$/6 UC FeTe bilayer, the PtIr tips are regularly conditioned on an MBE-grown Ag film to ensure clean and stable tunneling characteristics. The d$I$/d$V$ spectra are obtained using the standard lock-in method by applying an additional small a.c. excitation voltage at a frequency $f = 987.5$ Hz. All setpoints for our STM/S measurements are provided in the figure captions. Unless otherwise specified, all STM/S measurements are performed at $T = 310$ mK. The Lawler-Fujita drift-correction algorithm[65] is applied to all atomic-resolution images and spectroscopic imaging-STM measurements to eliminate drift artifacts. All STM data are treated using standard functions in MATLAB, Python 3.9, and WSxM 5.0 software[66].

**Josephson STM/S measurements**

In our Josephson STM/S measurements, polycrystalline superconducting Nb tips are used to form superconductor-insulator-superconductor (SIS) junctions. These superconducting Nb tips are made from 0.25 mm Nb wires via electrochemical etching in concentrated HCl solution using an a.c. voltage[67]. Before Josephson STM/S measurements, these Nb tips are first cleaned by electron-beam heating to remove surface oxides, and then gently conditioned on an MBE-grown Ag film to achieve a sharp apex of ~4 nm. The superconducting gap of our Nb tips ranges from 1.2 to 1.5 meV (Extended Data Fig. 5a)[68], with the variation likely influenced by the thickness of the Ag coating on the Nb tips. The $R_N$ values are ~1 MΩ for STS measurements on 1 QL $Sb_2Te_3$/6 UC FeTe (Fig. 3) and ~0.4 MΩ for STS measurements on 1 QL $Bi_2Te_3$/6 UC FeTe (Fig. 4). With $\Delta_{tip}$ = 1.37 meV and $\Delta_2$ = 3.60 meV for 1 QL $Sb_2Te_3$/6 UC FeTe and $\Delta_2$ = 3.32 meV for 1 QL $Bi_2Te_3$/6 UC FeTe, the Josephson coupling energy $E_J$ is estimated to be ~7.4 μeV and ~17.8 μeV, respectively (Supplementary Information). Both values are smaller than the thermal energy, $E_T = k_B T$ ~ 26.7 μeV at $T$ = 310 mK, confirming that our Josephson STM/S measurements are conducted in the phase-diffusive regime[34,36,37,57].

**Acknowledgments:** We thank P. J. Hirschfeld, C. Liu, L. Kong, Z. Y. Wang, P. Wu, and J. Yu for helpful discussions. The STM/S measurements are supported by the DOE grant (DE-SC0023113). The MBE growth is supported by the ONR Award (N000142412133) and the Penn State MRSEC for Nanoscale Science (DMR-2011839). WW acknowledges the support from the DOE grant (DE-SC0018153). Ziqiang W acknowledges the support from the DOE grant (DE-FG02-99ER45747).

XX and CZC acknowledge the support from the AFOSR grant (FA9550-21-1-0177). CZC acknowledges the support from the Gordon and Betty Moore Foundation's EPiQS Initiative (GBMF9063 to C.-Z. C).

**Author contributions:** CZC conceived and designed the experiment. Zihao W, BX, SP, ZJY, HR, and CZC performed the MBE growth. Zihao W, BX, SP, PX, JS, VG, and CZC performed all STM/S measurements. WW and XX provided experimental support. DX and Ziqiang W provided theoretical support. Zihao W and CZC analyzed the data and wrote the manuscript with input from all authors.

**Competing interests:** The authors declare no competing financial interests.

**Data availability:** The data that support the findings of this article are openly available[69].

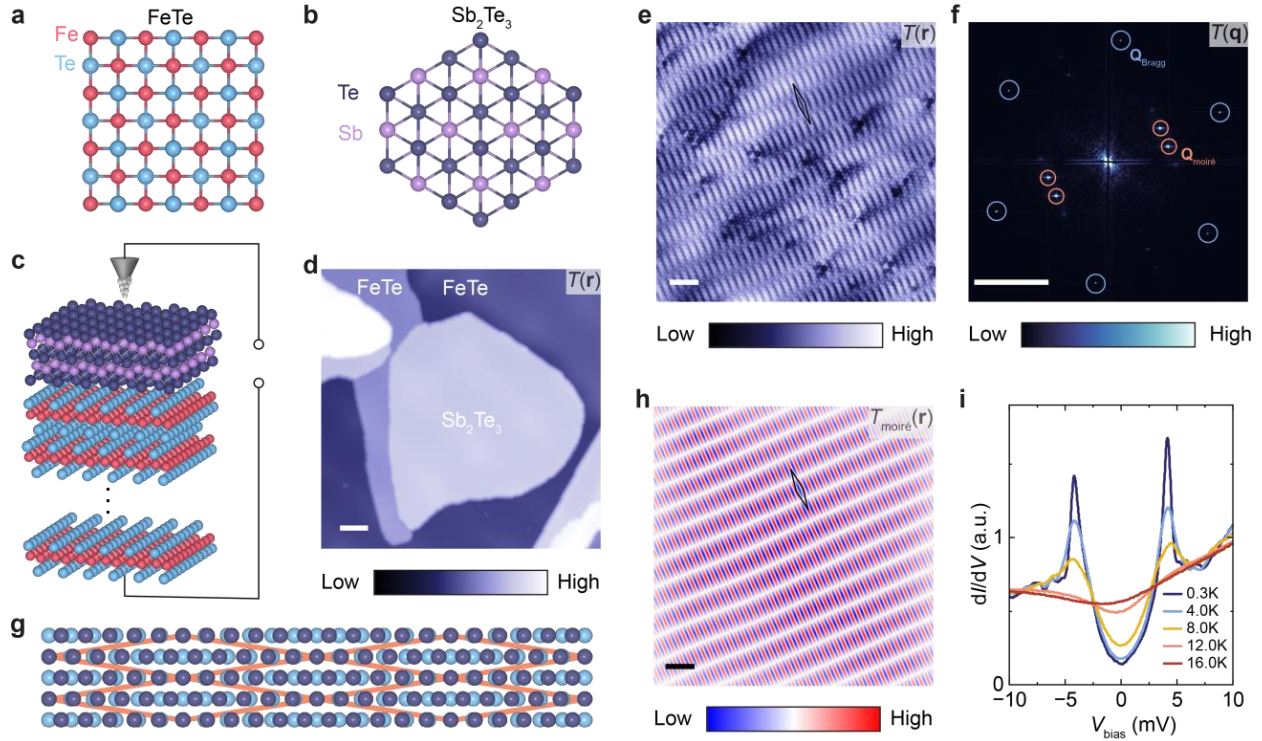

**Fig. 1| Moiré pattern and superconductivity in 1 QL Sb$_2$Te$_3$/6 UC FeTe bilayers. a**, **b**, Top views of the FeTe (**a**) and Sb$_2$Te$_3$ (**b**) lattice structures. **c**, Schematic of STM measurements on a Sb$_2$Te$_3$/FeTe bilayer. **d**, Large STM topography (100 × 100 nm$^2$) (setpoint bias $V_s$ = 2 V, setpoint current $I_s$ = 20 pA, and $T$ = 4.2 K). **e**, Atomic resolution STM image (30 × 30 nm$^2$) ($V_s$ = 50 mV and $I_s$ = 500 pA) of Sb$_2$Te$_3$ showing moiré superlattice. **f**, Fourier transform (FT) of (**e**). 6 Bragg peaks of the Te atoms in 1 QL Sb$_2$Te$_3$ and 4 moiré superlattice peaks are seen. **g**, Schematic of the moiré superlattice formed between the hexagonal Te lattice of Sb$_2$Te$_3$ and the square Te lattice of FeTe. **h**, FT-filtered image at 4 superlattice wavevectors $\mathbf{Q}_{moiré}$ in (**f**). The black rhombus in (**e**) and (**h**) is a moiré unit cell. **i**, Spatially averaged d$I$/d$V$ spectra at different $T$ ($V_s$ = 10 mV, $I_s$ = 1 nA, and excitation voltage $V_e$ = 0.1 mV). The superconducting gap closes at $T$ ~12 K. Scale bars: 10 nm (**d**); 3 nm (**e**); 1 Å$^{-1}$ (**f**); 3 nm (**h**).

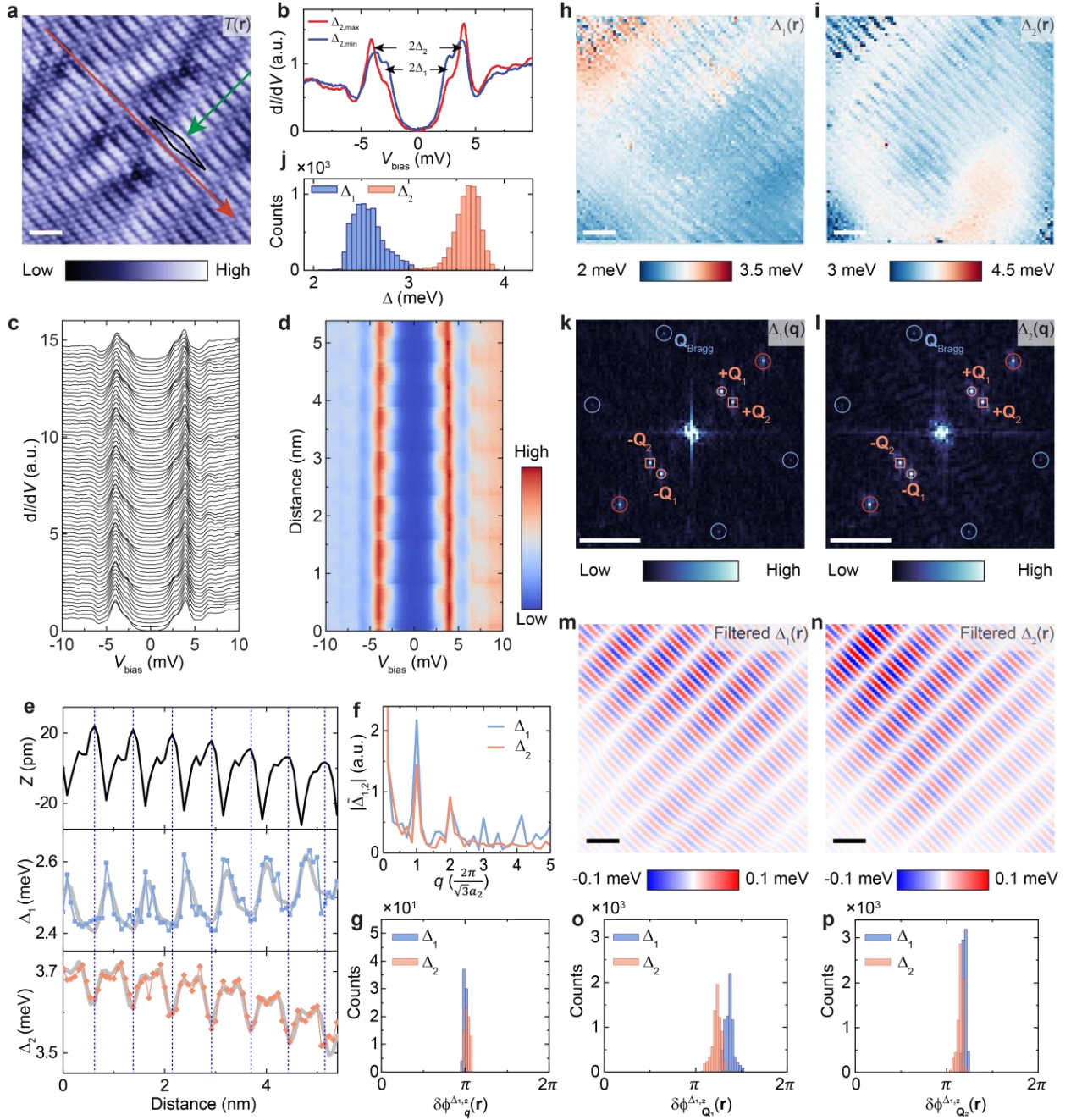

**Fig. 2| Moiré-induced CPDM states in 1 QL Sb$_2$Te$_3$/6 UC FeTe bilayers. a**, Atomic resolution STM image (14 × 14 nm$^2$) ($V_s$ = 50 mV and $I_s$ = 500 pA). The black rhombus is a moiré unit cell. **b**, Typical d$I$/d$V$ spectra at $\Delta_2$ maximum (red) and $\Delta_2$ minimum (blue). **c, d**, Waterfall plot (**c**) and color plot (**d**) of d$I$/d$V$ spectra measured along the green arrow in (**a**). All curves in (**c**) are vertically shifted for clarity. **e**, Extracted $Z$, $\Delta_1$, and $\Delta_2$ from the d$I$/d$V$ spectra in (**c**). The $Z$ value is simultaneously measured with the linecut d$I$/d$V$ spectra. The grey curves are 1D FT-filtered curves

of the fitting results. **f**, 1D FT curves of $\Delta_1$ and $\Delta_2$ in (**e**), showing two peaks at $q = 2\pi/(\sqrt{3}a_2)$ and $2q = 4\pi/(\sqrt{3}a_2)$. **g**, Distributions of relative phase changes between the moiré superlattice $\phi_q^Z$ and the superconducting gap size modulations $\phi_q^{\Delta_{1,2}}$. $\delta\phi_q^{\Delta_{1,2}} = \phi_q^{\Delta_{1,2}} - \phi_q^Z$. **h**, **i**, Spatially modulated superconducting gap maps of $\Delta_1$ (**h**) and $\Delta_2$ (**i**). Both $\Delta_1$ and $\Delta_2$ are modulated by the moiré pattern. **j**, Histogram of $\Delta_1$ (blue) and $\Delta_2$ (orange). The mean values of $\Delta_1$ and $\Delta_2$ are ~2.58 meV and ~3.60 meV, respectively. **k**, **l**, FT of $\Delta_1$ (**k**) and $\Delta_2$ (**l**) maps. **m**, **n**, FT-filtered images at $\pm Q_1$ and $\pm Q_2$ in (**k**, **l**). **o**, **p**, Distributions of relative phase changes between the moiré superlattice $\phi_Q^T$ and the superconducting gap size modulations $\phi_Q^{\Delta_{1,2}}$. $\delta\phi_{Q_1}^{\Delta_{1,2}} = \phi_{Q_1}^{\Delta_{1,2}} - \phi_{Q_1}^T$ (**o**) and $\delta\phi_{Q_2}^{\Delta_{1,2}} = \phi_{Q_2}^{\Delta_{1,2}} - \phi_{Q_2}^T$ (**p**). Scale bar: 2 nm (**a**, **h**, **i**, **m**, **n**); 1 Å$^{-1}$ (**k**, **l**). The STM setpoints in (**b-d**) and (**h**, **i**) are $V_s = 10$ mV, $I_s = 1$ nA, and $V_e = 0.1$ mV.

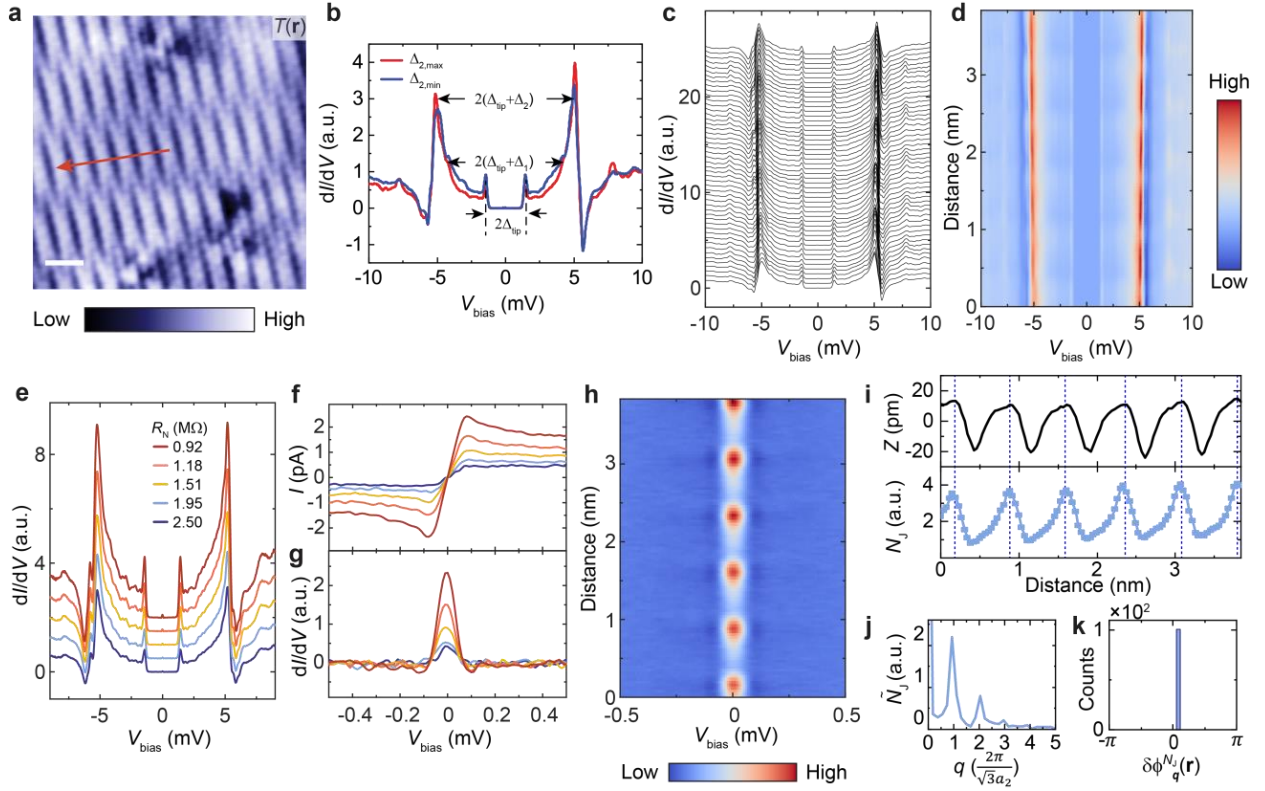

**Fig. 3| Real-space imaging of CPDM states in 1 QL Sb$_2$Te$_3$/6 UC FeTe bilayers. a**, Atomic resolution STM image (10 × 10 nm$^2$) ($V_s$ = 50 mV and $I_s$ = 1 nA). **b**, Typical d$I$/d$V$ spectra at $\Delta_2$ maximum (red) and $\Delta_2$ minimum (blue). Three pairs of d$I$/d$V$ peaks are observed at $\Delta_{tip}$, $\Delta_{tip} + \Delta_1$, and $\Delta_{tip} + \Delta_2$, where $\Delta_{tip}$ is the superconducting gap size of the Nb tip. **c, d**, Waterfall plot (**c**) and color plot (**d**) of d$I$/d$V$ spectra measured along the red arrow in (**a**). **e**, d$I$/d$V$ spectra measured at different $R_N$ by gradually reducing the tip-to-sample distance $D$ ($D_{offset}$ from 0 pm to 40 pm). Josephson tunneling signals are observed at $V_{bias}$ = 0 mV. **f, g**, $I$-$V_{bias}$ curves (top) and enlarged d$I$/d$V$ spectra near $V_{bias}$ = 0 mV (bottom). **h**, Color plot of $g_J(\mathbf{r}, V) \times R_N^2(\mathbf{r})$ spectra measured along the red arrow in (**a**). A coefficient $R_N^2$ is applied to d$I$/d$V$ to show the spatial modulation of the Cooper pair density. **i**, Extracted $Z$ (top) and $N_J(\mathbf{r})$ (bottom). **j**, 1D FT curve of $N_J(\mathbf{r})$ in (**h**). **k**, Distributions of relative phase changes between the moiré superlattice $\phi_q^Z$ and the superfluid density $\phi_q^{N_J}$. $\delta\phi_q^{N_J} = \phi_q^{N_J} - \phi_q^Z$. Scale bar: 1 nm (**a**). The STM setpoints in (**b-h**) are $V_s$ = 10 mV, $I_s$ = 4 nA, and $V_e$ = 0.05 mV. The value of $D_{offset}$ in (**h**) is 40 pm. All STM/S data are acquired using a superconducting Nb tip.

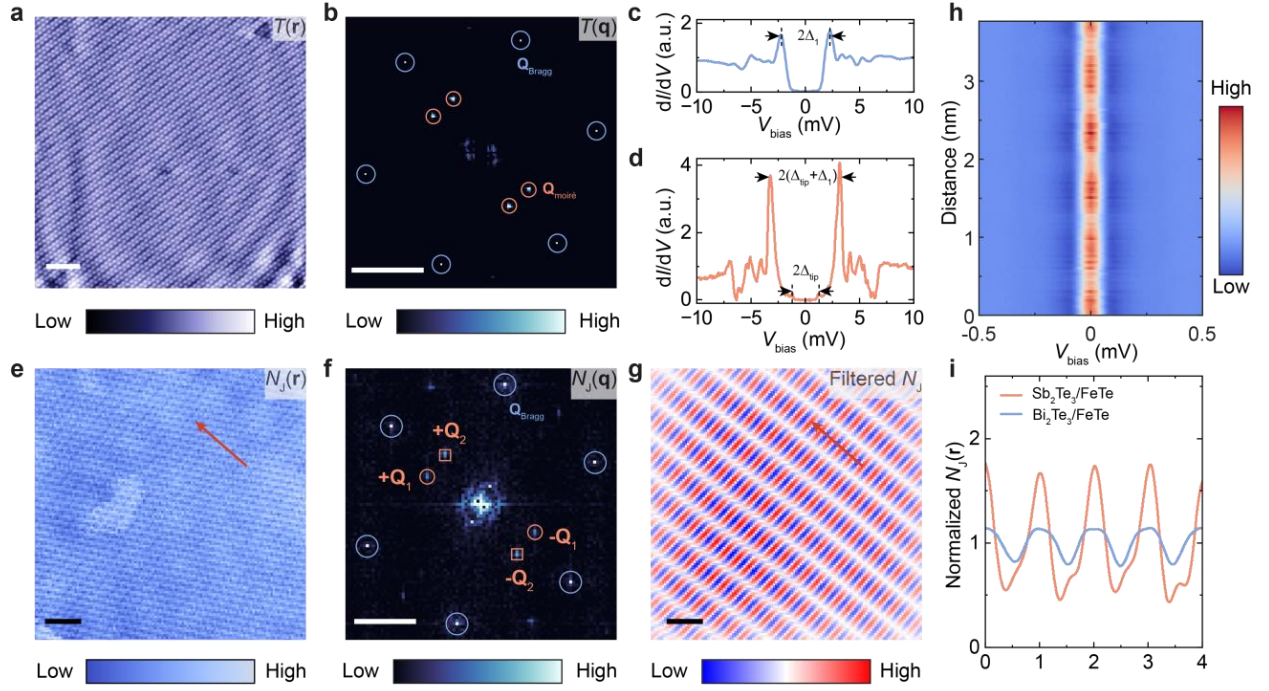

**Fig. 4| Moiré engineering of CPDM states in 1 QL (Bi,Sb)$_2$Te$_3$/6 UC FeTe bilayers. a**, Atomic resolution STM image (25 × 25 nm$^2$) ($V_s$ = -500 mV and $I_s$ = 4 nA). **b**, FT of (**a**). **c**, A typical d$I$/d$V$ spectrum measured by a PtIr tip ($V_s$ = 10 mV, $I_s$ = 500 pA, and $V_e$ = 0.2 mV). **d**, A typical d$I$/d$V$ spectrum measured by a superconducting Nb tip. **e**, $N_J(\mathbf{r})$ map measured by a superconducting Nb tip. **f**, FT of $N_J(\mathbf{r})$ in (**e**). The CPDM state conforms to the period of the moiré superlattice. **g**, FT-filtered $N_J(\mathbf{r})$ at ±$\mathbf{Q}_1$ and ±$\mathbf{Q}_2$ in (**f**). **h**, Color plot of $g_J(\mathbf{r}, V) \times R_N^2(\mathbf{r})$ spectra measured along the red arrows in (**e**) and (**g**). A coefficient $R_N^2$ is applied to d$I$/d$V$ to show the spatial modulation in Cooper pair density. **i**, Comparison of normalized $N_J(\mathbf{r})$ in Sb$_2$Te$_3$/FeTe and Bi$_2$Te$_3$/FeTe bilayers. Scale bars: 3 nm (**a**); 1 Å$^{-1}$ (**b**); 2 nm (**e**); 1 Å$^{-1}$ (**f**); 2 nm (**g**). The STM setpoints in (**d, e, h**) are $V_s$ = 10 mV, $I_s$ = 2 nA, and $V_e$ = 0.05 mV. The value of $D_{offset}$ in (**e, h**) is 125 pm.

**References**:

1. Kong, L., Papaj, M., Kim, H., Zhang, Y., Baum, E., Li, H., Watanabe, K., Taniguchi, T., Gu, G., Lee, P. A. & Nadj-Perge, S. Cooper-pair density modulation state in an iron-based superconductor. *Nature* **640**, 55-61 (2025).

2. Papaj, M., Kong, L., Nadj-Perge, S. & Lee, P. A. Pair density modulation from glide symmetry breaking and nematic superconductivity. *arXiv:2506.19903* (2025).

3. Wei, T., Liu, Y., Ren, W., Liang, Z., Wang, Z. & Wang, J. Observation of superconducting pair density modulation within lattice unit cell. *Chin. Phys. Lett.* **42**, 027404 (2025).

4. Cao, Y., Fatemi, V., Fang, S., Watanabe, K., Taniguchi, T., Kaxiras, E. & Jarillo-Herrero, P. Unconventional superconductivity in magic-angle graphene superlattices. *Nature* **556**, 43-50 (2018).

5. Cao, Y., Fatemi, V., Demir, A., Fang, S., Tomarken, S. L., Luo, J. Y., Sanchez-Yamagishi, J. D., Watanabe, K., Taniguchi, T., Kaxiras, E., Ashoori, R. C. & Jarillo-Herrero, P. Correlated insulator behaviour at half-filling in magic-angle graphene superlattices. *Nature* **556**, 80-84 (2018).

6. Yankowitz, M., Chen, S., Polshyn, H., Zhang, Y., Watanabe, K., Taniguchi, T., Graf, D., Young, A. F. & Dean, C. R. Tuning superconductivity in twisted bilayer graphene. *Science* **363**, 1059-1064 (2019).

7. Serlin, M., Tschirhart, C. L., Polshyn, H., Zhang, Y., Zhu, J., Watanabe, K., Taniguchi, T., Balents, L. & Young, A. F. Intrinsic quantized anomalous Hall effect in a moiré heterostructure. *Science* **367**, 900-903 (2020).

8. Oh, M., Nuckolls, K. P., Wong, D., Lee, R. L., Liu, X., Watanabe, K., Taniguchi, T. & Yazdani, A. Evidence for unconventional superconductivity in twisted bilayer graphene. *Nature* **600**, 240-245 (2021).

9. Guo, Y., Pack, J., Swann, J., Holtzman, L., Cothrine, M., Watanabe, K., Taniguchi, T., Mandrus, D. G., Barmak, K., Hone, J., Millis, A. J., Pasupathy, A. & Dean, C. R. Superconductivity in 5.0° twisted bilayer $WSe_2$. *Nature* **637**, 839-845 (2025).

10  Xia, Y., Han, Z., Watanabe, K., Taniguchi, T., Shan, J. & Mak, K. F. Superconductivity in twisted bilayer WSe$_2$. *Nature* **637**, 833-838 (2025).

11  Mao, J., Milovanovic, S. P., Andelkovic, M., Lai, X., Cao, Y., Watanabe, K., Taniguchi, T., Covaci, L., Peeters, F. M., Geim, A. K., Jiang, Y. & Andrei, E. Y. Evidence of flat bands and correlated states in buckled graphene superlattices. *Nature* **584**, 215-220 (2020).

12  Park, H., Cai, J., Anderson, E., Zhang, Y., Zhu, J., Liu, X., Wang, C., Holtzmann, W., Hu, C., Liu, Z., Taniguchi, T., Watanabe, K., Chu, J. H., Cao, T., Fu, L., Yao, W., Chang, C. Z., Cobden, D., Xiao, D. & Xu, X. Observation of fractionally quantized anomalous Hall effect. *Nature* **622**, 74-79 (2023).

13  Li, T., Jiang, S., Shen, B., Zhang, Y., Li, L., Tao, Z., Devakul, T., Watanabe, K., Taniguchi, T., Fu, L., Shan, J. & Mak, K. F. Quantum anomalous Hall effect from intertwined moiré bands. *Nature* **600**, 641-646 (2021).

14  Xu, F., Sun, Z., Jia, T. T., Liu, C., Xu, C., Li, C. S., Gu, Y., Watanabe, K., Taniguchi, T., Tong, B. B., Jia, J. F., Shi, Z. W., Jiang, S. W., Zhang, Y., Liu, X. X. & Li, T. X. Observation of integer and fractional quantum anomalous Hall effects in twisted bilayer MoTe$_2$. *Phys. Rev. X* **13**, 031037 (2023).

15  Li, H., Li, S., Regan, E. C., Wang, D., Zhao, W., Kahn, S., Yumigeta, K., Blei, M., Taniguchi, T., Watanabe, K., Tongay, S., Zettl, A., Crommie, M. F. & Wang, F. Imaging two-dimensional generalized Wigner crystals. *Nature* **597**, 650-654 (2021).

16  Tsui, Y. C., He, M., Hu, Y., Lake, E., Wang, T., Watanabe, K., Taniguchi, T., Zaletel, M. P. & Yazdani, A. Direct observation of a magnetic-field-induced Wigner crystal. *Nature* **628**, 287-292 (2024).

17  Li, H., Xiang, Z., Reddy, A. P., Devakul, T., Sailus, R., Banerjee, R., Taniguchi, T., Watanabe, K., Tongay, S., Zettl, A., Fu, L., Crommie, M. F. & Wang, F. Wigner molecular crystals from multielectron moiré artificial atoms. *Science* **385**, 86-91 (2024).

18  Thompson, E., Chu, K. T., Mesple, F., Zhang, X. W., Hu, C. W., Zhao, Y. Z., Park, H., Cai, J. Q., Anderson, E., Watanabe, K., Taniguchi, T., Yang, J. H., Chu, J. H., Xu, X. D., Cao, T.,

Xiao, D. & Yankowitz, M. Microscopic signatures of topology in twisted MoTe$_2$. *Nat. Phys.* **21**, 1224–1230 (2025).

19   Zhang, F., Morales-Durán, N., Li, Y. X., Yao, W., Su, J. J., Lin, Y. C., Dong, C. Y., Liu, X. H., Chen, F. X. R., Kim, H., Watanabe, K., Taniguchi, T., Li, X. Q., Robinson, J. A., Macdonald, A. H. & Shih, C. K. Experimental signature of layer skyrmions and implications for band topology in twisted WSe$_2$ bilayers. *Nat. Phys.* **21**, 1217–1223 (2025).

20   Naritsuka, M., Machida, T., Asano, S., Yanase, Y. & Hanaguri, T. Superconductivity controlled by twist angle in monolayer NbSe$_2$ on graphene. *Nat. Phys.* **21**, 746–753 (2025).

21   Fulde, P. & Ferrell, R. A. Superconductivity in a strong spin-exchange field. *Phys. Rev.* **135**, A550 (1964).

22   Larkin, A. I. Inhomogeneous state of superconductors. *Sov. Phys. JETP* **20**, 762 (1965).

23   Vershinin, M., Misra, S., Ono, S., Abe, Y., Ando, Y. & Yazdani, A. Local ordering in the pseudogap state of the high-$T_c$ superconductor Bi$_2$Sr$_2$CaCu$_2$O$_{8+\delta}$. *Science* **303**, 1995-1998 (2004).

24   Chen, H. D., Vafek, O., Yazdani, A. & Zhang, S. C. Pair density wave in the pseudogap state of high temperature superconductors. *Phys. Rev. Lett.* **93**, 187002 (2004).

25   Valla, T., Fedorov, A. V., Lee, J., Davis, J. C. & Gu, G. D. The ground state of the pseudogap in cuprate superconductors. *Science* **314**, 1914-1916 (2006).

26   Hamidian, M. H., Edkins, S. D., Kim, C. K., Davis, J. C., Mackenzie, A. P., Eisaki, H., Uchida, S., Lawler, M. J., Kim, E. A., Sachdev, S. & Fujita, K. Atomic-scale electronic structure of the cuprate *d*-symmetry form factor density wave state. *Nat. Phys.* **12**, 150-156 (2016).

27   Hamidian, M. H., Edkins, S. D., Joo, S. H., Kostin, A., Eisaki, H., Uchida, S., Lawler, M. J., Kim, E. A., Mackenzie, A. P., Fujita, K., Lee, J. & Davis, J. C. Detection of a Cooper-pair density wave in Bi$_2$Sr$_2$CaCu$_2$O$_{8+x}$. *Nature* **532**, 343-347 (2016).

28   Ruan, W., Li, X. T., Hu, C., Hao, Z. Q., Li, H. W., Cai, P., Zhou, X. J., Lee, D. H. & Wang, Y. Y. Visualization of the periodic modulation of Cooper pairing in a cuprate superconductor.

*Nat. Phys.* **14**, 1178-1182 (2018).

29  Edkins, S. D., Kostin, A., Fujita, K., Mackenzie, A. P., Eisaki, H., Uchida, S., Sachdev, S., Lawler, M. J., Kim, E. A., Seamus Davis, J. C. & Hamidian, M. H. Magnetic field–induced pair density wave state in the cuprate vortex halo. *Science* **364**, 976-980 (2019).

30  Du, Z., Li, H., Joo, S. H., Donoway, E. P., Lee, J., Davis, J. C. S., Gu, G., Johnson, P. D. & Fujita, K. Imaging the energy gap modulations of the cuprate pair-density-wave state. *Nature* **580**, 65-70 (2020).

31  Wang, S., Choubey, P., Chong, Y. X., Chen, W., Ren, W., Eisaki, H., Uchida, S., Hirschfeld, P. J. & Davis, J. C. S. Scattering interference signature of a pair density wave state in the cuprate pseudogap phase. *Nat. Commun.* **12**, 6087 (2021).

32  Liu, Y. Z., Wei, T. H., He, G. Y., Zhang, Y., Wang, Z. Q. & Wang, J. Pair density wave state in a monolayer high-$T_c$ iron-based superconductor. *Nature* **618**, 934-939 (2023).

33  Zhao, H., Blackwell, R., Thinel, M., Handa, T., Ishida, S., Zhu, X., Iyo, A., Eisaki, H., Pasupathy, A. N. & Fujita, K. Smectic pair-density-wave order in EuRbFe$_4$As$_4$. *Nature* **618**, 940-945 (2023).

34  Liu, X. L., Chong, Y. X., Sharma, R. & Davis, J. C. S. Discovery of a Cooper-pair density wave state in a transition-metal dichalcogenide. *Science* **372**, 1447-1452 (2021).

35  Cao, L., Xue, Y., Wang, Y., Zhang, F. C., Kang, J., Gao, H. J., Mao, J. & Jiang, Y. Directly visualizing nematic superconductivity driven by the pair density wave in NbSe$_2$. *Nat. Commun.* **15**, 7234 (2024).

36  Chen, H., Yang, H., Hu, B., Zhao, Z., Yuan, J., Xing, Y., Qian, G., Huang, Z., Li, G., Ye, Y., Ma, S., Ni, S., Zhang, H., Yin, Q., Gong, C., Tu, Z., Lei, H., Tan, H., Zhou, S., Shen, C., Dong, X., Yan, B., Wang, Z. & Gao, H. J. Roton pair density wave in a strong-coupling kagome superconductor. *Nature* **599**, 222-228 (2021).

37  Deng, H., Qin, H., Liu, G., Yang, T., Fu, R., Zhang, Z., Wu, X., Wang, Z., Shi, Y., Liu, J., Liu, H., Yan, X. Y., Song, W., Xu, X., Zhao, Y., Yi, M., Xu, G., Hohmann, H., Holbaek, S. C., Durrnagel, M., Zhou, S., Chang, G., Yao, Y., Wang, Q., Guguchia, Z., Neupert, T.,

Thomale, R., Fischer, M. H. & Yin, J. X. Chiral kagome superconductivity modulations with residual Fermi arcs. *Nature* **632**, 775-781 (2024).

38  Han, X., Chen, H., Tan, H., Cao, Z., Huang, Z., Ye, Y., Zhao, Z., Shen, C., Yang, H., Yan, B., Wang, Z. & Gao, H. J. Atomic manipulation of the emergent quasi-2D superconductivity and pair density wave in a kagome metal. *Nat. Nanotechnol.* **20**, 1017–1025 (2025).

39  Gu, Q., Carroll, J. P., Wang, S., Ran, S., Broyles, C., Siddiquee, H., Butch, N. P., Saha, S. R., Paglione, J., Davis, J. C. S. & Liu, X. Detection of a pair density wave state in UTe$_2$. *Nature* **618**, 921-927 (2023).

40  Zhang, Y., Yang, L., Liu, C., Zhang, W. & Fu, Y.-S. Visualizing uniform lattice-scale pair density wave in single-layer FeSe/SrTiO$_3$ films. *arXiv:2406.05693* (2024).

41  Koz, C., Rößler, S., Tsirlin, A. A., Wirth, S. & Schwarz, U. Low-temperature phase diagram of Fe$_{1+y}$Te studied using X-ray diffraction. *Phys. Rev. B* **88**, 094509 (2013).

42  Mansour, A. N., Wong-Ng, W., Huang, Q., Tang, W., Thompson, A. & Sharp, J. Structural characterization of Bi$_2$Te$_3$ and Sb$_2$Te$_3$ as a function of temperature using neutron powder diffraction and extended X-ray absorption fine structure techniques. *J. Appl. Phys.* **116**, 083513 (2014).

43  Liang, J., Zhang, Y. J., Yao, X., Li, H., Li, Z. X., Wang, J. N., Chen, Y. Z. & Sou, I. K. Studies on the origin of the interfacial superconductivity of Sb$_2$Te$_3$/Fe$_{1+y}$Te heterostructures. *Proc. Natl. Acad. Sci. USA* **117**, 221-227 (2020).

44  Yi, H., Hu, L. H., Zhao, Y. F., Zhou, L. J., Yan, Z. J., Zhang, R., Yuan, W., Wang, Z., Wang, K., Hickey, D. R., Richardella, A. R., Singleton, J., Winter, L. E., Wu, X., Chan, M. H. W., Samarth, N., Liu, C. X. & Chang, C. Z. Dirac-fermion-assisted interfacial superconductivity in epitaxial topological-insulator/iron-chalcogenide heterostructures. *Nat. Commun.* **14**, 7119 (2023).

45  He, Q. L., Liu, H., He, M., Lai, Y. H., He, H., Wang, G., Law, K. T., Lortz, R., Wang, J. & Sou, I. K. Two-dimensional superconductivity at the interface of a Bi$_2$Te$_3$/FeTe heterostructure. *Nat. Commun.* **5**, 4247 (2014).

46  Yi, H., Zhao, Y. F., Chan, Y. T., Cai, J., Mei, R., Wu, X., Yan, Z. J., Zhou, L. J., Zhang, R., Wang, Z., Paolini, S., Xiao, R., Wang, K., Richardella, A. R., Singleton, J., Winter, L. E., Prokscha, T., Salman, Z., Suter, A., Balakrishnan, P. P., Grutter, A. J., Chan, M. H. W., Samarth, N., Xu, X., Wu, W., Liu, C. X. & Chang, C. Z. Interface-induced superconductivity in magnetic topological insulators. *Science* **383**, 634-639 (2024).

47  Yuan, W., Yan, Z. J., Yi, H., Wang, Z., Paolini, S., Zhao, Y. F., Zhou, L., Wang, A. G., Wang, K., Prokscha, T., Salman, Z., Suter, A., Balakrishnan, P. P., Grutter, A. J., Winter, L. E., Singleton, J., Chan, M. H. W. & Chang, C. Z. Coexistence of superconductivity and antiferromagnetism in topological magnet $MnBi_2Te_4$ films. *Nano Lett.* **24**, 7962-7971 (2024).

48  Wang, Q. Y., Li, Z., Zhang, W. H., Zhang, Z. C., Zhang, J. S., Li, W., Ding, H., Ou, Y. B., Deng, P., Chang, K., Wen, J., Song, C. L., He, K., Jia, J. F., Ji, S. H., Wang, Y. Y., Wang, L. L., Chen, X., Ma, X. C. & Xue, Q. K. Interface-induced high-temperature superconductivity in single unit-cell FeSe films on $SrTiO_3$. *Chin. Phys. Lett.* **29**, 037402 (2012).

49  Yin, J. X., Wu, Z., Wang, J. H., Ye, Z. Y., Gong, J., Hou, X. Y., Shan, L., Li, A., Liang, X. J., Wu, X. X., Li, J., Ting, C. S., Wang, Z. Q., Hu, J. P., Hor, P. H., Ding, H. & Pan, S. H. Observation of a robust zero-energy bound state in iron-based superconductor Fe(Te, Se). *Nat. Phys.* **11**, 543-546 (2015).

50  Sprau, P. O., Kostin, A., Kreisel, A., Bohmer, A. E., Taufour, V., Canfield, P. C., Mukherjee, S., Hirschfeld, P. J., Andersen, B. M. & Davis, J. C. S. Discovery of orbital-selective Cooper pairing in FeSe. *Science* **357**, 75-80 (2017).

51  Chen, M., Chen, X., Yang, H., Du, Z., Zhu, X., Wang, E. & Wen, H. H. Discrete energy levels of Caroli-de Gennes-Matricon states in quantum limit in $FeTe_{0.55}Se_{0.45}$. *Nat. Commun.* **9**, 970 (2018).

52  Chen, M. Y., Chen, X. Y., Yang, H., Du, Z. Y. & Wen, H. H. Superconductivity with twofold symmetry in $Bi_2Te_3/FeTe_{0.55}Se_{0.45}$ heterostructures. *Sci. Adv.* **4**, eaat1084 (2018).

53  Wang, Z., Rodriguez, J. O., Jiao, L., Howard, S., Graham, M., Gu, G. D., Hughes, T. L., Morr, D. K. & Madhavan, V. Evidence for dispersing 1D Majorana channels in an iron-based

superconductor. *Science* **367**, 104-108 (2020).

54  Dynes, R. C., Narayanamurti, V. & Garno, J. P. Direct measurement of quasiparticle-lifetime broadening in a strong-coupled superconductor. *Phys. Rev. Lett.* **41**, 1509 (1978).

55  Du, Z., Yang, X., Lin, H., Fang, D., Du, G., Xing, J., Yang, H., Zhu, X. & Wen, H. H. Scrutinizing the double superconducting gaps and strong coupling pairing in $(Li_{1-x}Fe_x)OHFeSe$. *Nat. Commun.* **7**, 10565 (2016).

56  Fujita, K., Hamidian, M. H., Edkins, S. D., Kim, C. K., Kohsaka, Y., Azuma, M., Takano, M., Takagi, H., Eisaki, H., Uchida, S., Allais, A., Lawler, M. J., Kim, E. A., Sachdev, S. & Davis, J. C. Direct phase-sensitive identification of a *d*-form factor density wave in underdoped cuprates. *Proc. Natl. Acad. Sci. USA* **111**, E3026-E3032 (2014).

57  Cho, D., Bastiaans, K. M., Chatzopoulos, D., Gu, G. D. & Allan, M. P. A strongly inhomogeneous superfluid in an iron-based superconductor. *Nature* **571**, 541-545 (2019).

58  Kimura, H., Barber, R. P., Ono, S., Ando, Y. & Dynes, R. C. Scanning Josephson tunneling microscopy of single-crystal $Bi_2Sr_2CaCu_2O_{8+\delta}$ with a conventional superconducting tip. *Phys. Rev. Lett.* **101**, 037002 (2008).

59  Joo, S. H., Kim, J. J., Yoo, J. H., Park, M. S., Lee, K. S., Gu, G. & Lee, J. Cooper pair density of $Bi_2Sr_2CaCu_2O_{8+x}$ in atomic scale at 4.2 K. *Nano Lett.* **19**, 1112-1117 (2019).

60  Liu, Y. B., Zhou, J., Wu, C. & Yang, F. Charge-4e superconductivity and chiral metal in 45-twisted bilayer cuprates and related bilayers. *Nat. Commun.* **14**, 7926 (2023).

61  Li, Y. Y., Wang, G., Zhu, X. G., Liu, M. H., Ye, C., Chen, X., Wang, Y. Y., He, K., Wang, L. L. & Ma, X. C. Intrinsic topological insulator $Bi_2Te_3$ thin films on Si and their thickness limit. *Adv. Mater.* **22**, 4002-4007 (2010).

62  Jiang, Y., Wang, Y., Chen, M., Li, Z., Song, C., He, K., Wang, L., Chen, X., Ma, X. & Xue, Q.-K. Landau quantization and the thickness limit of topological insulator thin films of $Sb_2Te_3$. *Phys. Rev. Lett.* **108**, 016401 (2012).

63  Chang, C.-Z., Liu, C.-X. & MacDonald, A. H. Colloquium: Quantum anomalous hall effect. *Rev. Mod. Phys.* **95**, 011002 (2023).

64  Kezilebieke, S., Vano, V., Huda, M. N., Aapro, M., Ganguli, S. C., Liljeroth, P. & Lado, J. L. Moiré-enabled topological superconductivity. *Nano Lett.* **22**, 328-333 (2022).

65  Lawler, M. J., Fujita, K., Lee, J., Schmidt, A. R., Kohsaka, Y., Kim, C. K., Eisaki, H., Uchida, S., Davis, J. C., Sethna, J. P. & Kim, E. A. Intra-unit-cell electronic nematicity of the high-$T_c$ copper-oxide pseudogap states. *Nature* **466**, 347-351 (2010).

66  Horcas, I., Fernandez, R., Gomez-Rodriguez, J. M., Colchero, J., Gomez-Herrero, J. & Baro, A. M. WSXM: A software for scanning probe microscopy and a tool for nanotechnology. *Rev. Sci. Instrum.* **78**, 013705 (2007).

67  Uehara, Y., Fujita, T., Iwami, M. & Ushioda, S. Superconducting niobium tip for scanning tunneling microscope light emission spectroscopy. *Rev. Sci. Instrum.* **72**, 2097-2099 (2001).

68  Ternes, M., Schneider, W. D., Cuevas, J. C., Lutz, C. P., Hirjibehedin, C. F. & Heinrich, A. J. Subgap structure in asymmetric superconducting tunnel junctions. *Phys. Rev. B* **74**, 132501 (2006).

69  Wang, Z. Data for "Moiré Engineering of Cooper-Pair Density Modulation States". *Zenodo* (2025). https://doi.org/10.5281/zenodo.17139260

**Extended Data Figures:**

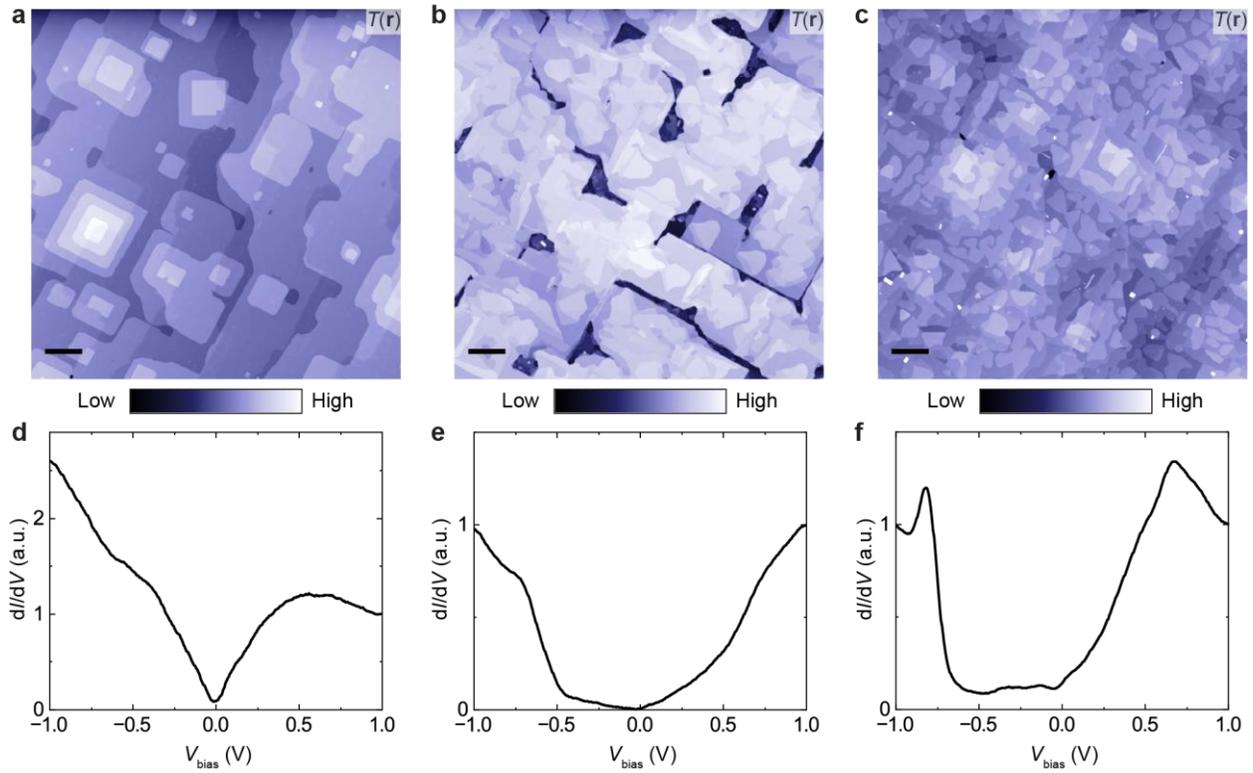

**Extended Data Fig. 1| 6 UC FeTe and 1 QL (Bi,Sb)$_2$Te$_3$/6 UC FeTe bilayers. a-c**, Large STM topography (1 × 1 μm$^2$) of 6 UC FeTe (**a**), 1 QL Sb$_2$Te$_3$/6 UC FeTe (**b**), and 1 QL Bi$_2$Te$_3$/6 UC FeTe (**c**). **d-f**, Average d$I$/d$V$ spectra on 6 UC FeTe (**d**), 1 QL Sb$_2$Te$_3$/6 UC FeTe (**e**), and 1 QL Bi$_2$Te$_3$/6 UC FeTe (**f**). Scale bars: 100 nm (**a-c**). STM setpoints: $V_s$ = 1.5 V and $I_s$ = 50 pA (**a, c**); $V_s$ = 2 V and $I_s$ = 50 pA (**b**); $V_s$ = 1 V, $I_s$ = 1 nA, and $V_e$ = 20 mV (**d, e**); $V_s$ = 1 V, $I_s$ = 500 pA, and $V_e$ = 20 mV(**f**).

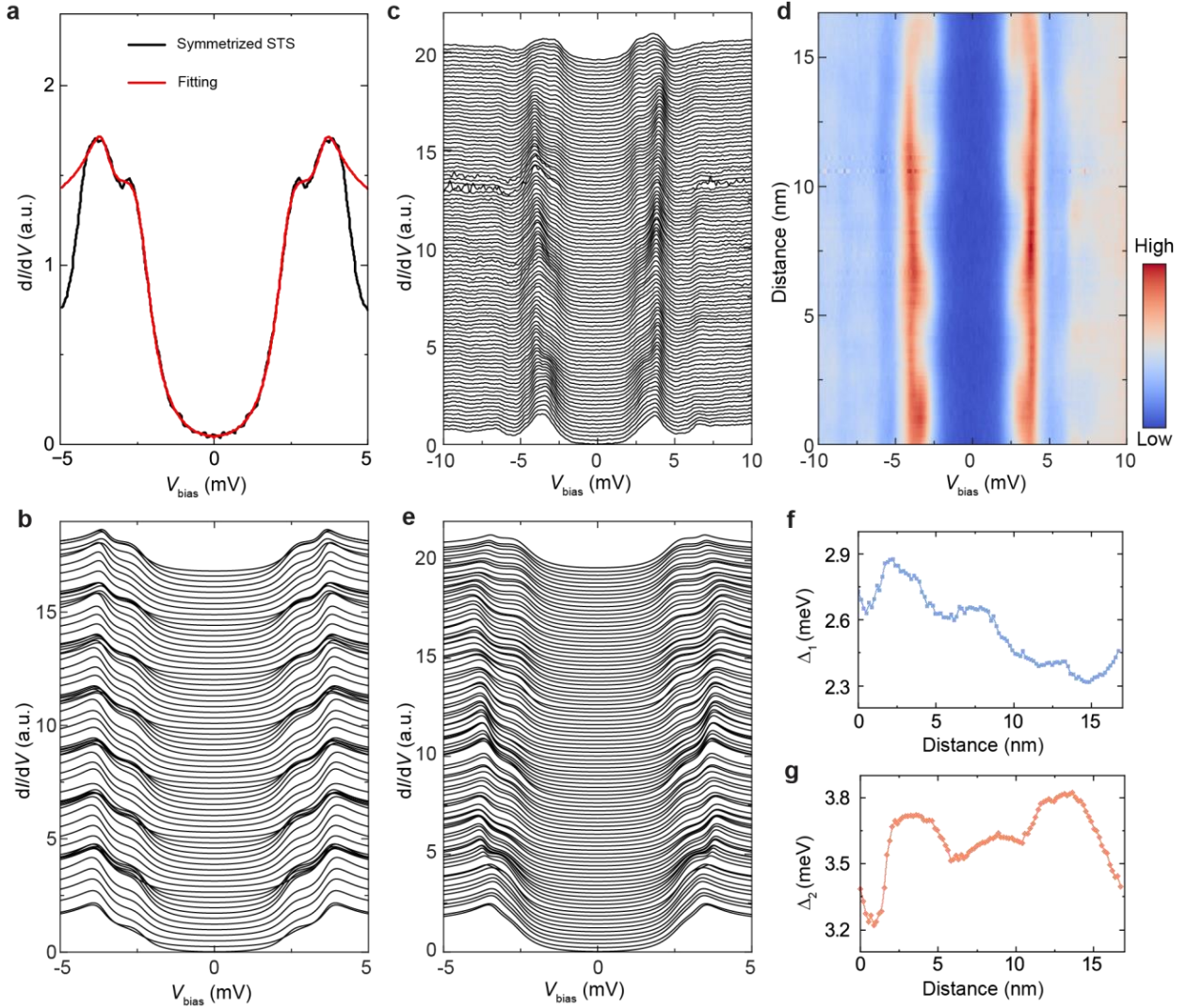

**Extended Data Fig. 2| Dynes model fits with two isotropic *s*-wave superconducting gaps. a**, Symmetrized d*I*/d*V* spectrum (black) and two-gap Dynes model fit (red). **b**, Two-gap Dynes model fits of the d*I*/d*V* spectra in Fig. 2c. **c**, **d**, Waterfall plot (**c**) and color plot (**d**) of d*I*/d*V* spectra measured along the major diagonal direction (red arrow, Fig. 2a) of the rhombic moiré pattern. **e**, Two-gap Dynes model fits of the d*I*/d*V* spectra in (**c**). **f**, **g**, Extracted $\Delta_1$ (**f**) and $\Delta_2$ (**g**) from (**e**). Both $\Delta_1$ and $\Delta_2$ exhibit modulations along the major diagonal direction of the rhombic moiré pattern.

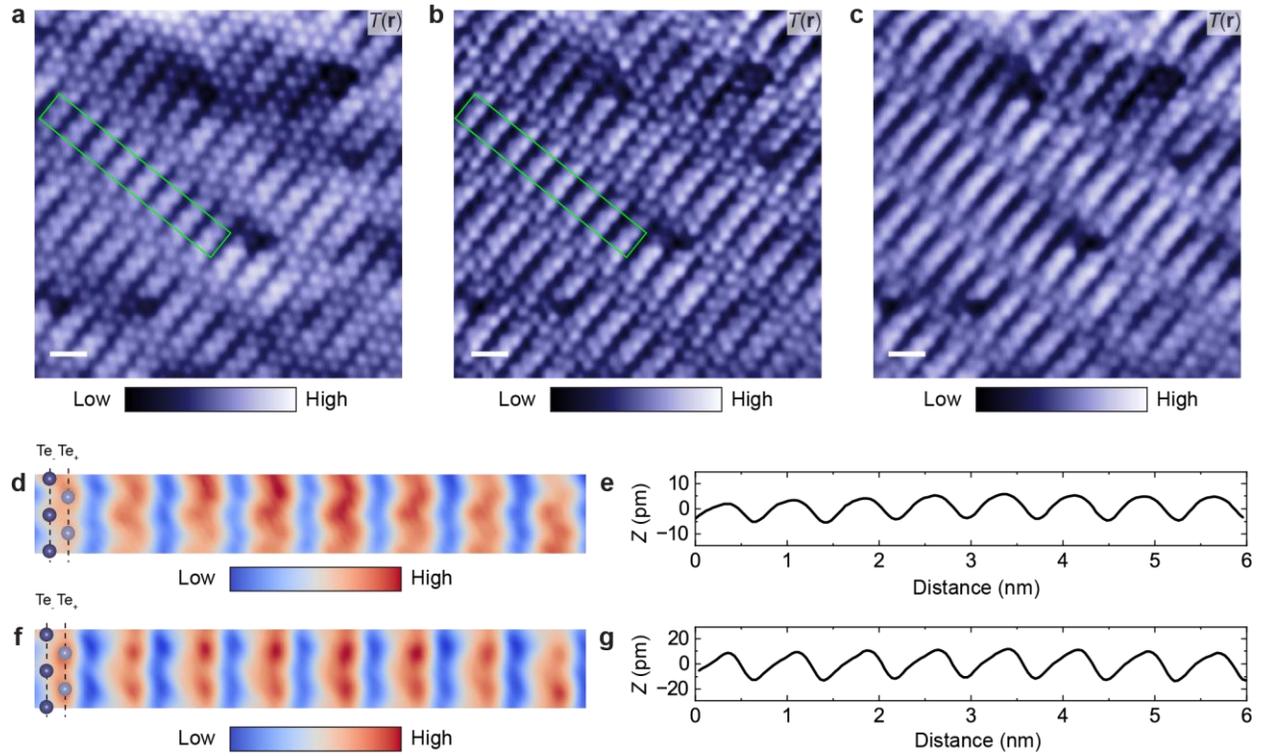

**Extended Data Fig. 3| Moiré pattern in 1 QL $Sb_2Te_3$/6 UC FeTe bilayers under different biases**. **a-c**, Atomic resolution STM images (10 × 10 nm$^2$) measured at $V_s$ = 50 mV (**a**), 10 mV (**b**), and 4 mV (**c**). **d, f**, Enlarged atomic resolution STM images of the area marked by the green rectangles in (**a**) and (**b**). $Te_+$ (light purple) and $Te_-$ (dark purple) atoms are the same at $V_s$ = 50 mV, while $Te_+$ atoms are higher than $Te_-$ atoms at $V_s$ = 10 mV. **e, g**, Vertically averaged height profile Z of (**d**) and (**f**). Z at $V_s$ = 10 mV reveals the difference between $Te_+$ and $Te_-$ atoms. Scale bars: 1 nm (**a-c**). STM setpoints: $I_s$ = 500 pA (**a**); $I_s$ = 500 pA (**b**); $I_s$ = 50 pA (**c**).

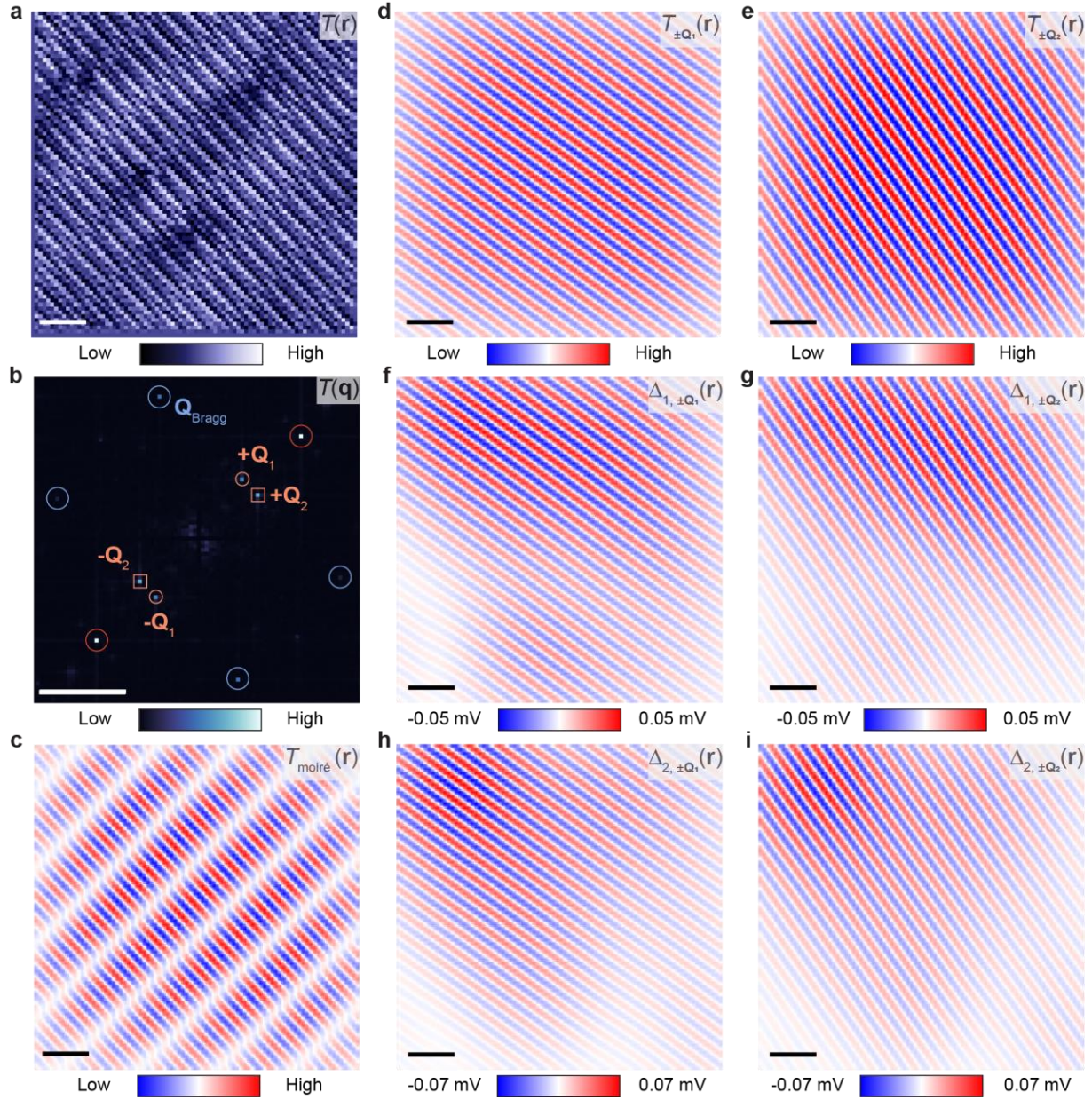

**Extended Data Fig. 4| FT-filtered images of the moiré superlattice, Δ$_1$, and Δ$_2$ in 1 QL Sb$_2$Te$_3$/6 UC FeTe bilayers. a**, Atomic resolution STM image (14 × 14 nm$^2$) acquired simultaneously with the spectroscopic imaging-STM measurements in Fig. 2h,i. **b**, FT of (**a**). **c-e**, FT-filtered images of the moiré superlattice at ±**Q**$_1$ (**d**) and ±**Q**$_2$ (**e**), and their sum (**c**). **f, g**, FT-filtered images of Δ$_1$ at ±**Q**$_1$ (**f**) and ±**Q**$_2$ (**g**). **h, i**, FT-filtered images of Δ$_2$ at ±**Q**$_1$ (**h**) and ±**Q**$_2$ (**i**). The gap modulations in (**f-i**) exhibit phase shifts relative to the moiré superlattice in (**d, e**), which can also be extracted using the 2D lock-in method (Figs. 2o,p, and S3). Scale bars: 2 nm (**a, c-i**); 1 Å$^{-1}$ (**b**).

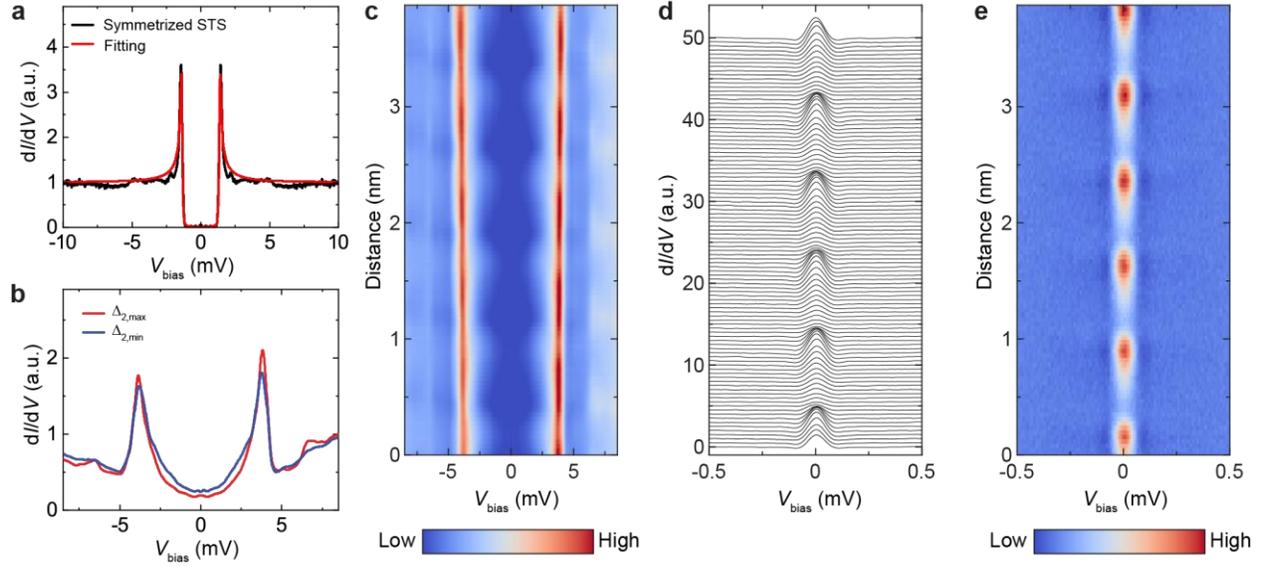

**Extended Data Fig. 5| Deconvoluted d$I$/d$V$ spectra and Josephson tunneling signals in 1 QL Sb$_2$Te$_3$/6 UC FeTe bilayers. a**, Superconducting gap of the Nb tip. The red curve is the Dynes model fit. The superconducting gap size of the Nb tip $\Delta_{tip}$ = 1.37 meV. **b, c,** d$I$/d$V$ spectra of 1 QL Sb$_2$Te$_3$/6 UC FeTe in Fig. 3b (**b**) and Fig. 3d (**c**) after numerically deconvolving the Nb tip contribution. The superconducting gap modulation in (**c**) is consistent with Fig. 2d. **d, e,** Waterfall plot (**d**) and color plot (**e**) of $g(\mathbf{r}, V)$ spectra measured along the red arrow in Fig. 3a.

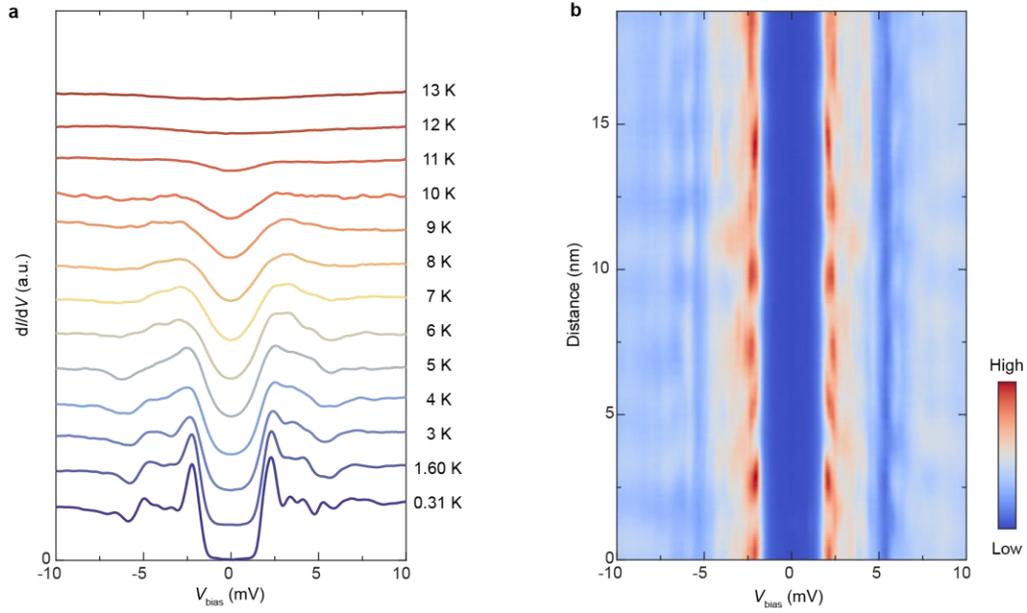

**Extended Data Fig. 6| More d$I$/d$V$ spectra of the 1 QL Bi$_2$Te$_3$/6 UC FeTe bilayer. a,** Typical d$I$/d$V$ spectra measured at different $T$ ($V_s$ = 10 mV, $I_s$ = 500 pA, and $V_e$ = 0.2 mV). The d$I$/d$V$ spectra are vertically shifted for clarity. **b,** Color plot of d$I$/d$V$ spectra measured along a 19 nm line ($V_s$ = 10 mV, $I_s$ = 2 nA, and $V_e$ = 0.05 mV). A clear spatial variation is observed in the spectra outside the superconducting gap $\Delta_1$ ~ 1.96 meV.

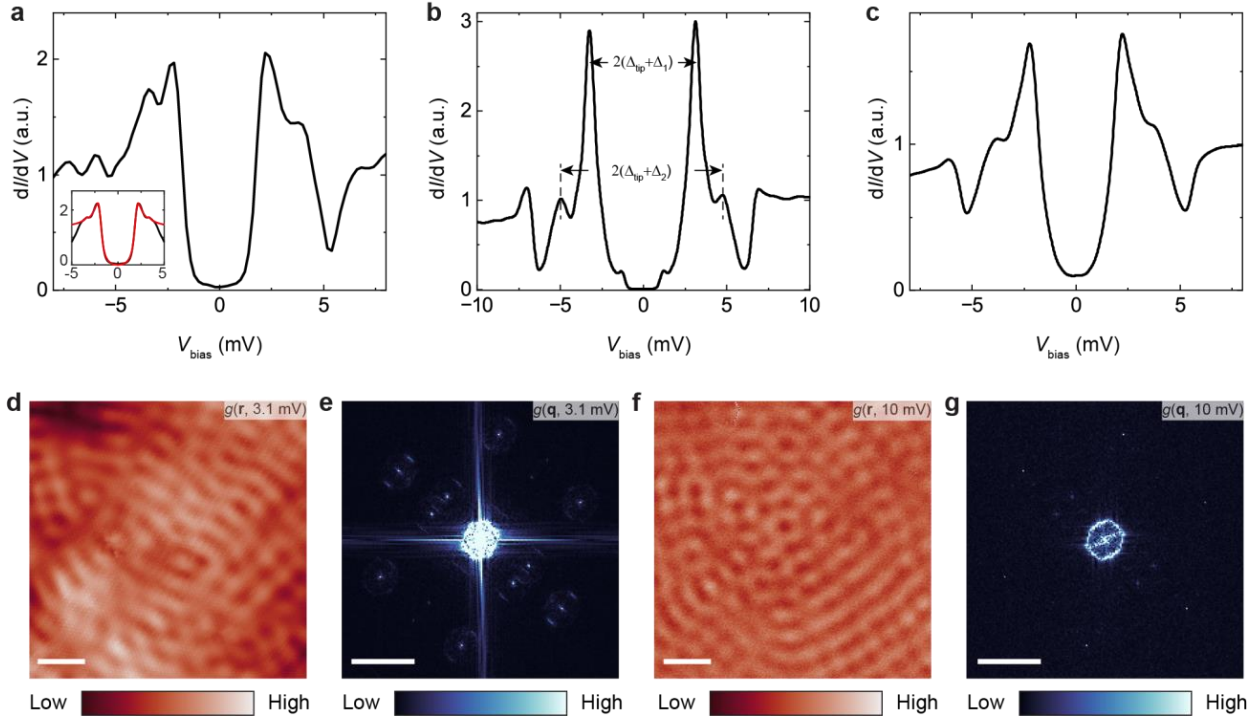

**Extended Data Fig. 7| Spatially averaged d$I$/d$V$ spectra of the 1 QL Bi$_2$Te$_3$/6 UC FeTe bilayer. a,** Spatially averaged d$I$/d$V$ spectrum from spectroscopic imaging-STM measurements using a PtIr tip ($V_s$ = 8 mV, $I_s$ = 1 nA, and $V_e$ = 0.2 mV) in Fig. S8. The average d$I$/d$V$ spectra remove contributions of QPI and the moiré pattern, revealing two pairs of coherence peaks. Inset: Dynes model fit of the d$I$/d$V$ spectrum. The black curve shows the symmetrized spectrum, and the red curve shows the fit, yielding $\Delta_1$ ~ 1.96 meV and $\Delta_2$ ~ 3.32 meV. **b,** Spatially averaged d$I$/d$V$ spectrum over a 170 × 170 grid in a 30 × 30 nm$^2$ area from spectroscopic imaging-STM measurements using a superconducting Nb tip ($V_s$ = 10 mV, $I_s$ = 1 nA, and $V_e$ = 0.2 mV). **c,** d$I$/d$V$ spectrum in (**b**) after numerically deconvolving the Nb tip contribution. The positions of the coherence peaks match those in (**a**) measured by a PtIr tip. **d,** Differential conductance map measured at convoluted coherence peak $E$ ~ $\Delta_1$+ $\Delta_{tip}$ using a Nb tip ($V_s$ = 3.1 mV, $I_s$ = 100 pA, and $V_e$ = 0.2 mV). **e,** FT of (**d**). **f,** Differential conductance map measured outside the convoluted superconducting gap using a Nb tip ($V_s$ = 10 mV, $I_s$ = 500 pA, and $V_e$ = 0.2 mV). **g,** FT of (**f**). FT of the QPI patterns in (**e**) and (**g**) exhibit a hexagonal shape, consistent with a prior study on 2 QL Bi$_2$Te$_3$/Fe(Se,Te) (ref.[52]). Scale bars: 5 nm (**d, f**); 1 Å$^{-1}$ (**e, g**).

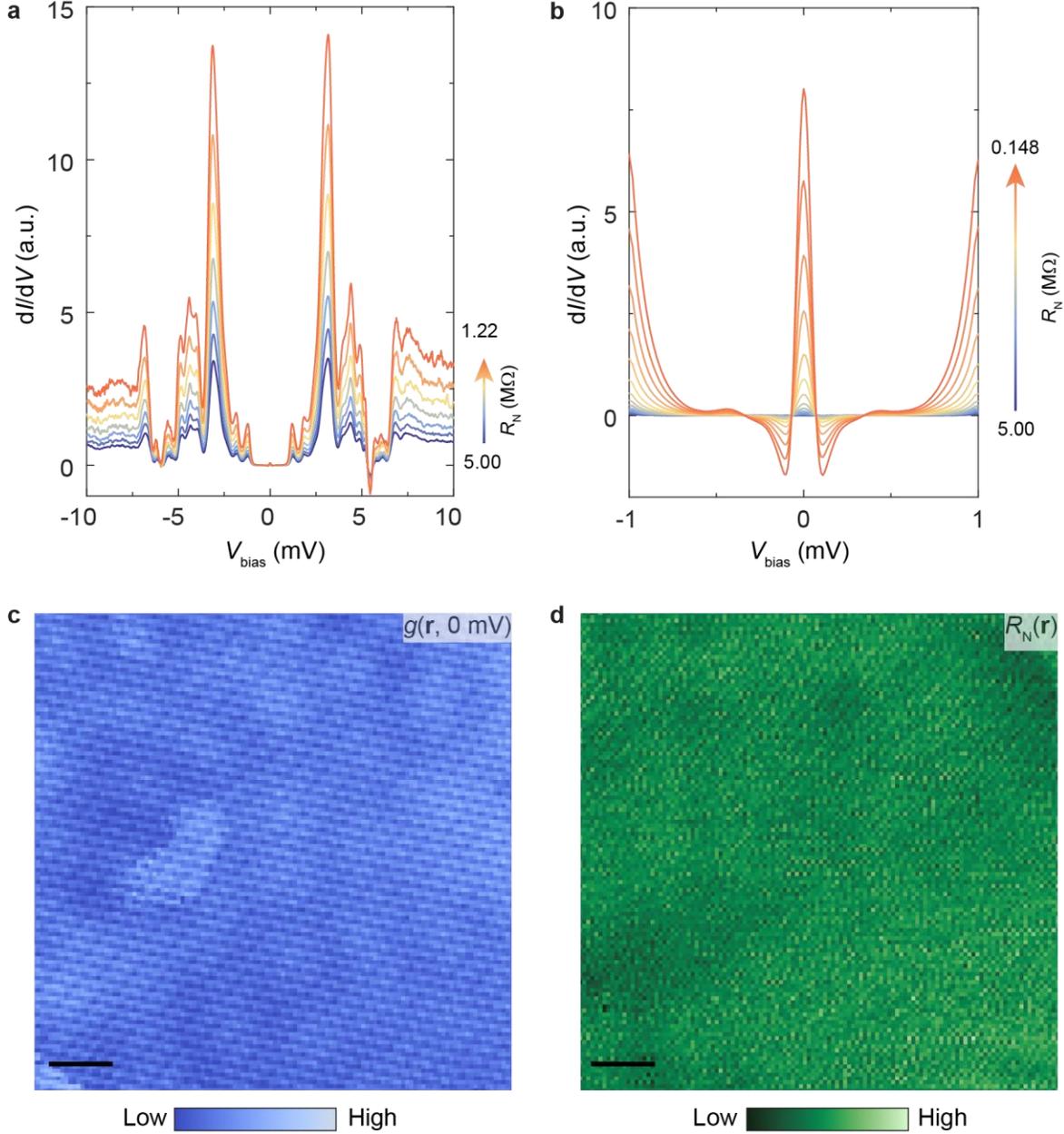

**Extended Data Fig. 8| Josephson STM/S measurements on 1 QL Bi$_2$Te$_3$/6 UC FeTe bilayers. a,** d$I$/d$V$ spectra measured at different $R_N$ by gradually reducing the tip-to-sample distance $D$ ($D_{\text{offset}}$ from 0 pm to 60 pm). **b,** d$I$/d$V$ spectra near $V_{\text{bias}}$ = 0 mV at different $R_N$ ($D_{\text{offset}}$ from 0 pm to 150 pm). **c,** Zero-bias d$I$/d$V$ maps showing Josephson tunneling signals. **d,** $R_N(\mathbf{r})$ map measured in the same area as (**c**). $N_J(\mathbf{r})$ map in Fig. 4e is obtained by calculating $g_J(\mathbf{r}, V=0\text{ mV}) \times R_N^2(\mathbf{r})$. Scale bars: 2 nm (**c, d**). The STM setpoints in (**a-c**) are $V_s$ = 10 mV, $I_s$ = 2 nA, and $V_e$ = 0.05 mV. The value of $D_{\text{offset}}$ in (**c**) is 125 pm.

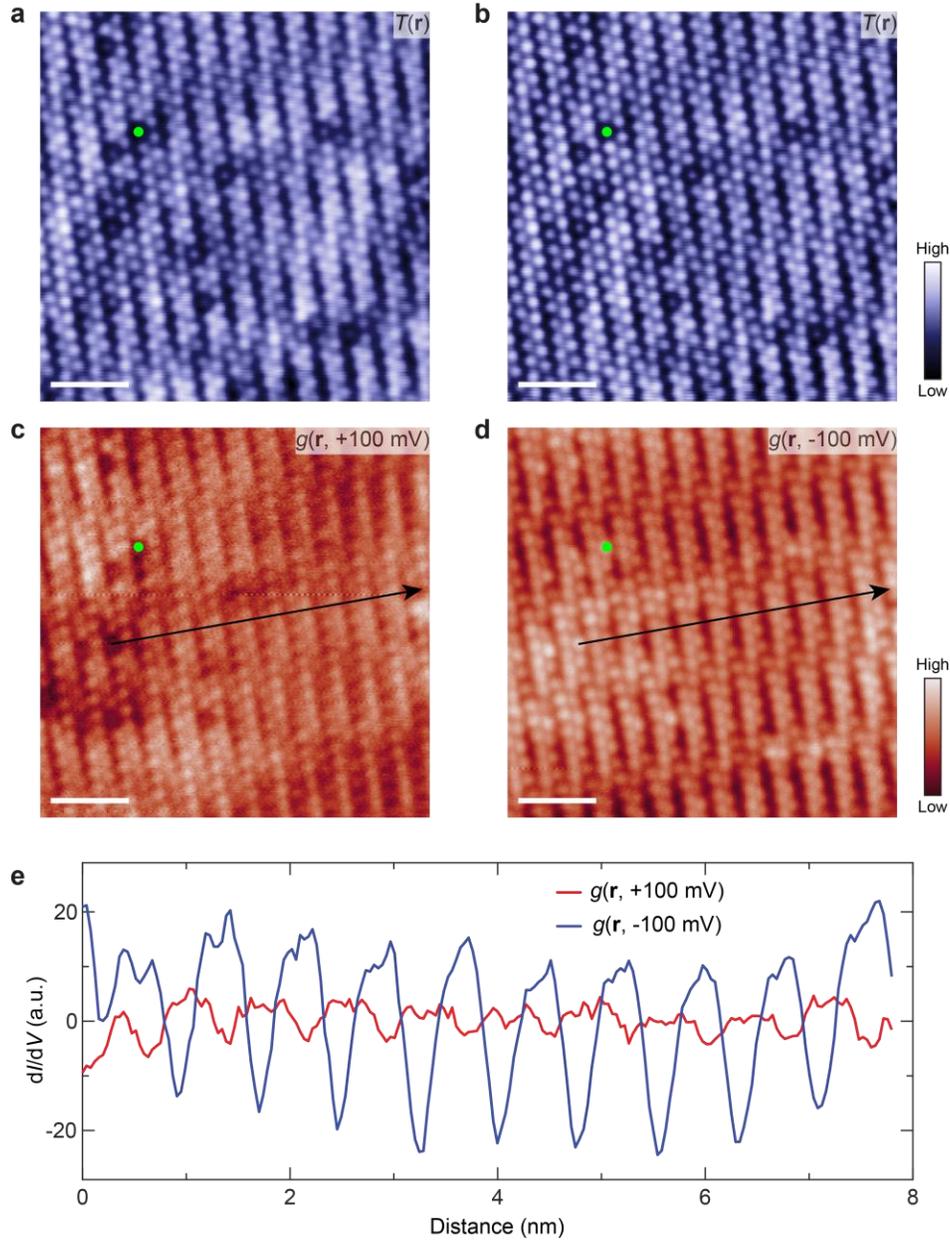

**Extended Data Fig. 9| d$I$/d$V$ maps of the 1 QL Sb$_2$Te$_3$/6 UC FeTe bilayer measured with opposite bias polarities. a**, Atomic resolution STM image (10 × 10 nm$^2$) measured at $V_s$ = 100 mV and $I_s$ = 2 nA. **b**, Atomic resolution STM image (10 × 10 nm$^2$) measured at $V_s$ = -100 mV and $I_s$ = 2 nA. **c**, $g(\mathbf{r}, 100\ \text{mV})$ obtained simultaneously with (**a**). **d**, $g(\mathbf{r}, -100\ \text{mV})$ obtained simultaneously with (**b**). The green dots in (**a-d**) mark the same defect, serving as a spatial reference. **e**, Line profile of the black curve in (**c, d**), demonstrating contrast inversion in the differential conductance. The excitation voltage $V_e$ in (**a-d**) is 10 mV.

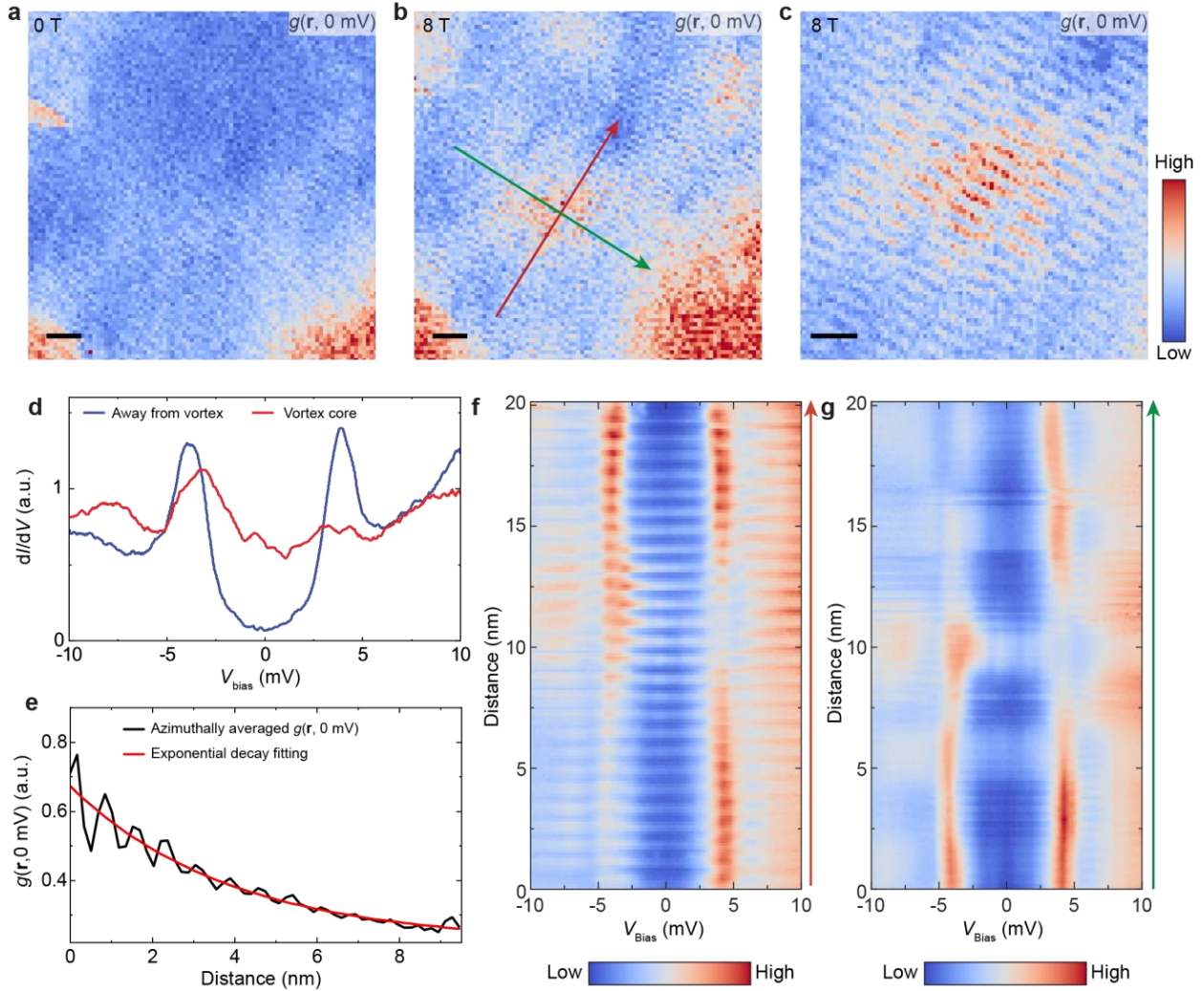

**Extended Data Fig. 10| Abrikosov vortex in 1 QL Sb$_2$Te$_3$/6 UC FeTe bilayers. a, b,** Zero-bias d$I$/d$V$ maps $g(\mathbf{r}, V=0\text{ mV})$ at $\mu_0 H = 0$ T (**a**) and 8 T (**b**). **c,** Zero-bias d$I$/d$V$ maps $g(\mathbf{r}, V=0\text{ mV})$ of a single Abrikosov vortex. **d,** d$I$/d$V$ spectra measured at $\mu_0 H = 8$ T at the vortex core (red) and away from the vortex (blue). **e,** Azimuthally averaged zero-bias d$I$/d$V$ $g(\mathbf{r}, V=0\text{ mV})$ around the vortex in (**c**) (black) and the exponential decay fit (red). The coherence length is estimated to be $\xi = 3.8 \pm 0.5$ nm. **f,** d$I$/d$V$ spectra along the red arrow in (**b**) (i.e., the minor diagonal direction of the rhombic moiré pattern). **g,** d$I$/d$V$ spectra along the green arrow in (**b**) (i.e., the major diagonal direction of the rhombic moiré pattern). The coherence peaks are suppressed near the vortex core, and $g(\mathbf{r}, V=0\text{ mV})$ exhibits spatial modulations, indicating a moiré-modulated normal state. Scale bars: 3 nm (**a, b**); 2 nm (**c**). STM setpoints: $V_s = 10$ mV, $I_s = 500$ pA, and $V_e = 0.2$ mV (**a-c**); $V_s = 10$ mV, $I_s = 1$ nA, and $V_e = 0.1$ mV (**d, f, g**).